\documentclass[11pt]{article}

\usepackage{graphicx}

\usepackage[left=1in, right=1in, top=1in, bottom=1in, includehead, includefoot, foot=0pt, head=0pt]{geometry} 
\parskip=\medskipamount 
\usepackage{amssymb} 
\usepackage{amsmath} 
\usepackage{amsbsy} 
\usepackage{booktabs} 
\usepackage{multirow} 
\usepackage[scientific-notation=false, group-separator={,}]{siunitx} 
\usepackage[small,bf,hang]{caption}	
\usepackage[round, colon, authoryear]{natbib}
\usepackage{hyperref}
\hypersetup{
	colorlinks=true,
	citecolor=[RGB]{17,110,138},
	linkcolor=[RGB]{17,110,138},
	urlcolor=[RGB]{17,110,138},
}
\usepackage{titling} 
\usepackage{todonotes}

\usepackage{pifont}
\usepackage{authblk}

\usepackage{float}
\usepackage{subcaption}

\usepackage{eurosym}

\usepackage{bbm}

\usepackage{lipsum}

\usepackage{fancyhdr}
\usepackage{changepage}

\usepackage{mathrsfs}

\usepackage{outlines}

\usepackage{soul}
\newcommand{\ulcolor}[2][Red]{\setulcolor{#1}\ul{#2}}

\makeatletter
\newcommand*\bigcdot{\mathpalette\bigcdot@{.5}}
\newcommand*\bigcdot@[2]{\mathbin{\vcenter{\hbox{\scalebox{#2}{$\m@th#1\bullet$}}}}}
\makeatother

\usepackage[noend]{algorithm2e} 
\SetAlCapFnt{\small} 
\SetAlCapNameFnt{\small} 
\usepackage{setspace} 

\usepackage{tikz} 
\usetikzlibrary{matrix, fit, decorations.pathreplacing} 

\usetikzlibrary{arrows.meta} 
\usepackage[outline]{contour} 
\contourlength{1.4pt}

\usetikzlibrary{trees,shapes,decorations, automata, arrows}

\tikzstyle{data} = [rectangle, rounded corners, minimum width=4.5cm, minimum height=1cm,text centered, text width = 4.5cm, draw=black, fill=orange!30]
\tikzstyle{process} = [rectangle, minimum width=4.5cm, minimum height=1cm, text centered, draw=black, text width = 4.5cm, fill=violet!30]
\tikzstyle{decisiont} = [diamond, minimum width=3cm, minimum height=3cm, text centered, text width=3cm, draw=black, fill=orange!30]
\tikzstyle{arrow} = [thick,->,>=stealth]
\tikzstyle{datae} = [rectangle, rounded corners, minimum width=4.5cm, minimum height=1cm,text centered, text width = 4.5cm, draw=black, fill=green!30]
\tikzstyle{processt} = [rectangle, minimum width=4.5cm, minimum height=1cm, text centered, draw=black, text width = 4.5cm, fill=orange!30]

\usepackage{adjustbox}

\usepackage{nicematrix}

\usepackage{tablefootnote}

\usepackage{diagbox}
\newcolumntype{P}[1]{>{\centering\arraybackslash} m{#1}} 

\usepackage{tabularx}
\makeatletter
\def\hlinewd#1{%
\noalign{\ifnum0=`}\fi\hrule \@height #1 %
\futurelet\reserved@a\@xhline}
\makeatother
\newcolumntype{?}[1]{!{\vrule width #1}}

\usepackage{array}

\usepackage{arydshln}

\usepackage{makecell}

\usepackage{colortbl}

\definecolor{myRed}{RGB}{228,15,15}
\definecolor{myGreen}{RGB}{40,150,40}
\definecolor{myOrange}{RGB}{255, 150, 0}
\definecolor{myPink}{RGB}{255, 20, 147}
\definecolor{myPurple}{RGB}{204,20, 167}
\definecolor{kul-secblue}	{RGB}{220,231,240}	
\xdefinecolor{UUxblue}{cmyk}{0.806,0.139,0,0.576}
\definecolor{myRed}{RGB}{255, 40, 40}

\definecolor{kul-blue}		{RGB}{29,141,176}		
\definecolor{kul-blue2}{RGB}{17,110,138} 			
\definecolor{kul-secblue}	{RGB}{220,231,240}		
\definecolor{kul-lightblue} {RGB}{82,189,236}
\definecolor{kul-dark}		{RGB}{47,77,93}			
\definecolor{gray}			{gray}{.5}				
\definecolor{lgray}			{gray}{.9}				

\definecolor{tabwhite}{HTML}{FFFFFF}
\definecolor{tabgray}{HTML}{FAFAFF}

\usepackage{calc}
\graphicspath{{graphics/}}

\usepackage{svg}

\usepackage[many]{tcolorbox}
\newtcolorbox{rightbrace}[1]{%
    enhanced jigsaw, 
    frame hidden, 
    overlay={%
        \draw [
            fill=none,
            gray,
        ];
        \node at ([shift={(2mm,0mm)}]frame.east) [anchor=center, gray, rotate=270] {#1}; 
    },
    parbox=false,
}

\usepackage[round,colon,authoryear]{natbib}

\setcitestyle{aysep={}}

\usepackage[defaultcolor=black, authormarkup=none]{changes}
\definechangesauthor[name=roundtwo, color=blue]{roundtwo}

\begin{document}

\title{\huge A multi-view contrastive learning framework for spatial embeddings in risk modelling}
\author[1,2]{Freek Holvoet}
\author[3]{Christopher Blier-Wong}
\author[1,2,4]{Katrien Antonio}
\affil[1]{Faculty of Economics and Business, KU Leuven, Belgium.}
\affil[2]{LRisk, Leuven Research Center on Insurance and Financial Risk Analysis, KU Leuven, Belgium.}
\affil[3]{Department of Statistical Sciences, University of Toronto, Canada}
\affil[4]{Faculty of Economics and Business, University of Amsterdam, The Netherlands.}
\date{\today}

{\centering \maketitle}
\thispagestyle{empty}

\begin{abstract}
\noindent Incorporating spatial information, particularly when related to climate, weather, and demographic factors, is crucial for improving underwriting precision and enhancing risk management in insurance. However, spatial data are often unstructured, high-dimensional, and difficult to integrate into predictive models. Embedding methods are needed to convert spatial data into meaningful representations for modelling tasks. We propose a novel multi-view contrastive learning framework for generating spatial embeddings that combine information from multiple spatial data sources. To train the model, we construct a spatial dataset that merges satellite imagery and OpenStreetMap features across Europe. The framework aligns these spatial views with coordinate-based encodings, producing low-dimensional embeddings that capture both spatial structure and contextual similarity. Once trained, the model generates embeddings directly from latitude–longitude pairs, enabling any dataset with coordinates to be enriched with meaningful spatial features without requiring access to the original spatial inputs. In a case study on French real estate prices, we compare models trained on raw coordinates against those using our spatial embeddings as inputs. The embeddings consistently improve predictive accuracy across generalised linear, additive, and boosting models, while providing \added{post-hoc explainable spatial effects and demonstrating generalisation of the fitted spatial effects to regions without training observations}. \added{A second case study on flood claim counts across Belgian postal codes confirms that the embeddings improve territorial risk classification in an insurance context.}
\\[.5\baselineskip]
\end{abstract}

\section{Introduction}


Spatial data are central to fields such as environmental monitoring, urban planning, and real estate pricing. Spatial modelling is also increasingly important in insurance applications. For instance, spatial data are used to link weather patterns to hail-damage claim frequency \citep{GAO2022161} or to quantify the effect of heat waves on mortality rates \citep{Robben_EnvMort}. These examples illustrate how incorporating spatial information enables actuaries to better analyse and manage risks that exhibit spatial dependence. Despite their importance, integrating spatial information into predictive models remains challenging. Location is often represented as a simple coordinate pair (latitude and longitude) or as a categorical variable, which fails to capture real-world geographical structure. Popular machine learning methods, such as random forests and gradient boosting machines, apply linear splits on these coordinate variables, leading to artificial rectangular patterns that do not align with real-world geography \citep{Geerts2024-dk}.


Early work in spatial risk modelling was primarily inspired by geostatistics, where the focus is on understanding the dependence structures inherent in spatially distributed data. Two modelling strategies are typically explored: incorporating location via spatial smoothing of residuals, see \citet{taylor1989use, boskov1994premium, taylor2001geographic} or by considering the spatial coordinates directly in a statistical model, for example, in a generalised additive model with a bivariate smooth effect of latitude and longitude, see \citet{dimakos2002bayesian, gschlossl2007spatial, shi2017territorial, Henckaerts_binning, henckaerts2020boosting}. While these approaches model spatial dependence explicitly, modern machine learning increasingly represents spatial information through representation learning, where spatial information is captured implicitly via so-called embeddings.


Embeddings are vector representations of data objects, obtained through an encoder that maps discrete or structured inputs into a continuous, typically low-dimensional space. The dimensionality of the embedding determines how much information it can represent: low-dimensional embeddings capture the most essential structure of the input data, while higher-dimensional embeddings can preserve more detailed or complex patterns. A particular class of embeddings, called entity embeddings, refers to compact, continuous-valued representations of categorical variables, where each category is mapped to a low-dimensional vector \citep{GuoCheng2016EEoC}. The use of embeddings in actuarial science is not new: \citet{richman2020ai} discusses supervised entity embeddings for categorical variables, and \citet{blier-wong2021rethinking} construct entity embeddings via unsupervised learning. In a motor third-party liability dataset, categorical variables such as gender, vehicle type, and coverage level can each be represented by entity embeddings. Applications in non-life insurance pricing include \citet{delong2021, avanzi2024machine, richman2024highcardinality, wilsens2024reducing, holvoet2023neural}. Spatial information can be categorical in nature, for instance, when postal codes or administrative regions are treated as levels of a categorical variable. In such cases, entity embeddings can be used to capture spatial variation in risk levels, as shown in \citet{shi2022nonlife}. In \citet{blier-wong2022geographic}, the authors claim that entity embeddings are inappropriate for integrating spatial information in insurance loss modelling. Granular territories such as postal codes are typically created to facilitate mail delivery, not to capture homogeneous risk profiles. Furthermore, if neighbours live on the border between two territories, their insurance premiums may differ since the embeddings of their territories will vary, even though their spatial risk should be similar. For this reason, the effect of spatial information in prediction models should be smooth on a map, and entity embeddings typically do not possess this property. 


For these reasons, \citet{blier-wong2022geographic} propose modelling spatial risk through spatial embeddings rather than entity embeddings. Spatial embeddings represent geographic information directly in a continuous latent space, where locations that are close in space are assigned similar vectors. Unlike entity embeddings, which treat each spatial unit as an independent category, spatial embeddings encode smooth transitions across space.

Spatial embeddings can also incorporate external contextual information about each location, such as census statistics, land cover maps, or satellite imagery. These context-aware spatial embeddings enable models to represent not only the location itself, but also what is present there. In a case study on claim frequency modelling using Canadian data, \citet{blier-wong2022geographic} construct spatial embeddings from geographic coordinates enriched with census data using a convolutional neural network. They show that these embeddings capture spatial risk patterns more effectively than traditional spatial smoothing methods, such as generalised additive models with bivariate smooth effects. Similarly, \citet{moussaid2023convolutional} constructs spatial embeddings based on H3 hexagonal coordinates \citep{uber_h3}, augmenting them with spatial context from OpenStreetMap data for each hexagonal cell. Using a Belgian motor third-party liability claims dataset, they demonstrate that incorporating such contextual information improves predictive accuracy. \added{These studies provide direct evidence that context-aware spatial embeddings are an effective tool for actuarial risk modelling, with \citet{blier-wong2022geographic} demonstrating their value on home insurance pricing for a Canadian portfolio. The present paper builds on this evidence and focuses on the construction of the embedding model itself, by proposing a framework that combines multiple spatial data sources within a single contrastive learning architecture.}


In the broader field of geospatial artificial intelligence (GeoAI), spatial representation learning has advanced rapidly through neural networks that process large-scale, heterogeneous spatial data \citep{yin2019gps2vec, blier-wong2020encoding, mai2020multiscale, mai2022review}. These methods typically produce task-agnostic spatial embeddings, which means that the embeddings are not trained for a single predictive task, but rather to capture general spatial structure and relationships across different regions or data sources. Once learnt, these embeddings can serve as general-purpose spatial representations that can be reused in different predictive settings. This property makes them particularly suited for \added{the generalisation of fitted spatial effects to territories where no outcome observations are available. By leveraging shared geographic characteristics through spatial embeddings, a downstream model can produce meaningful predictions in locations without training data.}


The recent development of foundation models extends context-aware spatial embedding models by aiming to create general-purpose geospatial representations that integrate multiple data modalities while retaining spatial awareness. Foundation models have already transformed language and vision through large pretrained architectures, and similar ambitions are now emerging for multimodal spatial models that unify imagery, vector data, and semantic context \citep{mai2024opportunities, mai2025next}. A recent example is AlphaEarth \citep{brown2025alphaearthfoundationsembeddingfield}, a large-scale multi-view and spatiotemporal geospatial foundation model that integrates satellite imagery, weather reanalysis, and auxiliary contextual data to produce globally consistent embeddings. One specific spatial foundational model architecture we will investigate in this paper is the SatCLIP model, proposed by \cite{klemmer2023towards, klemmer2023satclip}. SatCLIP constructs context-aware spatial embeddings by combining geographic coordinates with visual context extracted from satellite imagery. The model produces two types of embeddings: one derived from spatial coordinates using spherical harmonics to capture global spatial structure \citep{Ruswurm2023}, and another derived from satellite images using a vision transformer that extracts high-level visual patterns \citep{tarasiou2023vits}. These two embedding spaces are aligned through contrastive learning, where the model learns to assign similar representations to spatial locations that share similar visual and geographic characteristics, and dissimilar representations to locations that differ. The contrastive training procedure allows the coordinate encoder to internalise the contextual information contained in the imagery. As a result, SatCLIP can generate embeddings for new locations using only their geographic coordinates, without requiring access to the original spatial data. \added{For this reason we build on the SatCLIP architecture. In actuarial applications a dataset typically records the location of a policyholder or insured object but no further spatial context, so useful and rich embeddings must be generated from coordinates alone. Alternative context-aware approaches in the actuarial literature, such as \citet{blier-wong2022geographic} and \citet{moussaid2023convolutional}, construct embeddings directly from the contextual data and therefore require access to these data sources whenever embeddings are produced for new locations.} \citet{klemmer2023satclip} uses the trained SatCLIP model to capture large-scale spatial variation across the entire earth’s surface resulting in a high-dimensional embedding. They demonstrate the use of these embeddings in a case study on global temperature pattern prediction.

\begin{figure}[ht!] 
    \centering 
    \includegraphics[width=0.9\linewidth]{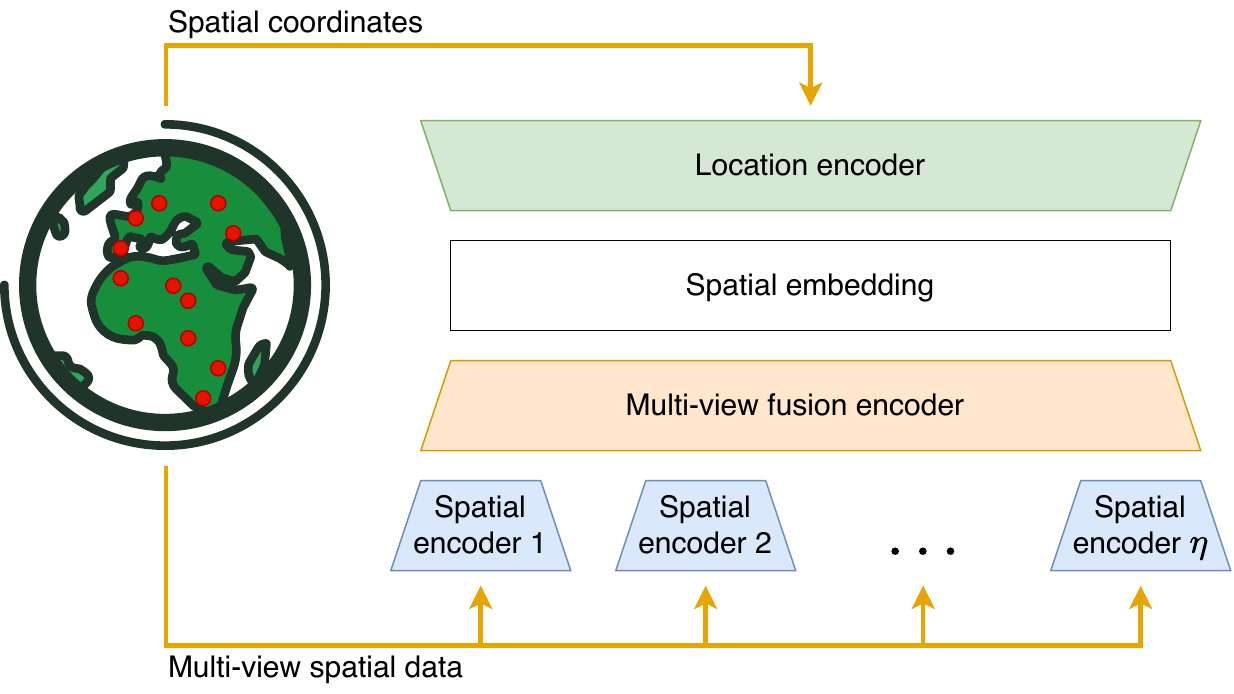} 
    \caption{\added{Schematic overview of our multi-view spatial embedding model. At the bottom, each blue block encodes one spatial view, and the orange block fuses these into a single spatial view embedding. At the top, the green block encodes the latitude and longitude coordinates. The model is trained with a contrastive learning objective, which aligns the coordinate embedding of a location with the embedding of its spatial views. Image inspired by \citet{klemmer2023satclip}.}} 
    \label{fig:multiviewmodel} 
\end{figure}

In this paper, we propose a multi-view contrastive learning model that generates spatial embeddings by leveraging multiple input data sources as spatial context. Our contribution lies in combining information from different spatial views with geographical coordinates to obtain low-dimensional embeddings that capture local spatial patterns. \added{This contribution is methodological in nature and applies to spatial representation learning in general. The design choices, however, are guided by actuarial practice, where datasets record the location of policyholders or insured objects and spatial features must be available from coordinates alone. To the best of our knowledge, an architecture that fuses multiple spatial views with a coordinate encoder through contrastive learning has not yet been proposed in the actuarial literature.} The model takes as input different sources of spatial data, such as satellite imagery and tabular neighbourhood information, together with the geographical coordinates of where the spatial information was gathered. Through contrastive learning, the model learns to associate patterns in the geographic coordinates with the information contained in the spatial views. Rather than targeting global coverage, the proposed framework is intended for regional settings. It can be retrained efficiently on consumer-grade hardware, flexibly incorporates different spatial data sources, and yields fine-grained spatial embeddings that capture local geographic structure in greater detail when compared to global-scale spatial embedding models. Our work differs from SatCLIP \citep{klemmer2023towards, klemmer2023satclip} in two key aspects. First, we integrate multiple spatial data sources rather than relying on a single modality, such as satellite imagery. Second, our model is designed to learn compact embeddings optimised for fine-grained regional applications, whereas SatCLIP produces high-dimensional global representations tailored to large-scale earth system modelling. \added{Figure \ref{fig:multiviewmodel} gives a schematic overview of the framework, with all details deferred to Section \ref{sec:model_archi}. The model consists of two parts. The first part, shown at the bottom of the figure, summarises the available spatial data for a location, in our case a satellite image and neighbourhood information from OpenStreetMap, into a single vector. The blue blocks encode each spatial view and the orange block fuses these into one vector. The second part, shown at the top, converts the latitude and longitude coordinates of that location into a vector of the same dimension, represented by the green block. During training, the model compares the two vectors and adjusts both parts so that, for each location, the coordinate-based vector matches the data-based vector, while differing from the vectors of other locations. After training, the green block on its own produces the spatial embedding. The embedding of any new location can therefore be computed from its coordinates, without requiring access to the underlying spatial data.}

This paper is structured as follows. Section \ref{sec:spatialdata} details the construction of a multi-view spatial dataset for European countries. We gather both satellite images via Google Maps \citep{google_maps_static_api} and neighbourhood context via OpenStreetMap \citep{openstreetmap} for almost $100\,000$ locations across Europe. Section \ref{sec:model_archi} shows the construction of the multi-view spatial embedding model and details the contrastive loss function, the training, and the hyperparameter choices. In Section \ref{sec:frpricing}, we illustrate the practical value of our spatial embeddings in \added{two case studies. The case studies serve as a demonstration and validation of this methodology, rather than as fully developed pricing applications. The first case study considers} the context of real estate pricing. Traditional models in this field often rely on basic location indicators, such as postal codes or proximity to landmarks, which only provide limited spatial context. In contrast, our proposed spatial embeddings incorporate a richer set of geographic features, enabling the model to better account for the influence of surrounding areas and environmental characteristics on property values. To evaluate our approach, we compare the predictive performance for predicting real estate prices using our spatial embeddings versus raw coordinates. We use permutation-based variable importance and partial dependence effects to examine how the embeddings influence the predictions, and \added{evaluate how the fitted spatial effects generalise} to areas held out of training. \added{The second case study, in Section \ref{sec:flood_casestudy}, applies the embeddings in an insurance setting and models flood claim counts across Belgian postal codes.} Section \ref{sec:conclusion} concludes the paper.

\section{Dataset of locations and spatial views}\label{sec:spatialdata}

\begin{figure}[ht!]
    \centering
    \includegraphics[width=0.5\linewidth]{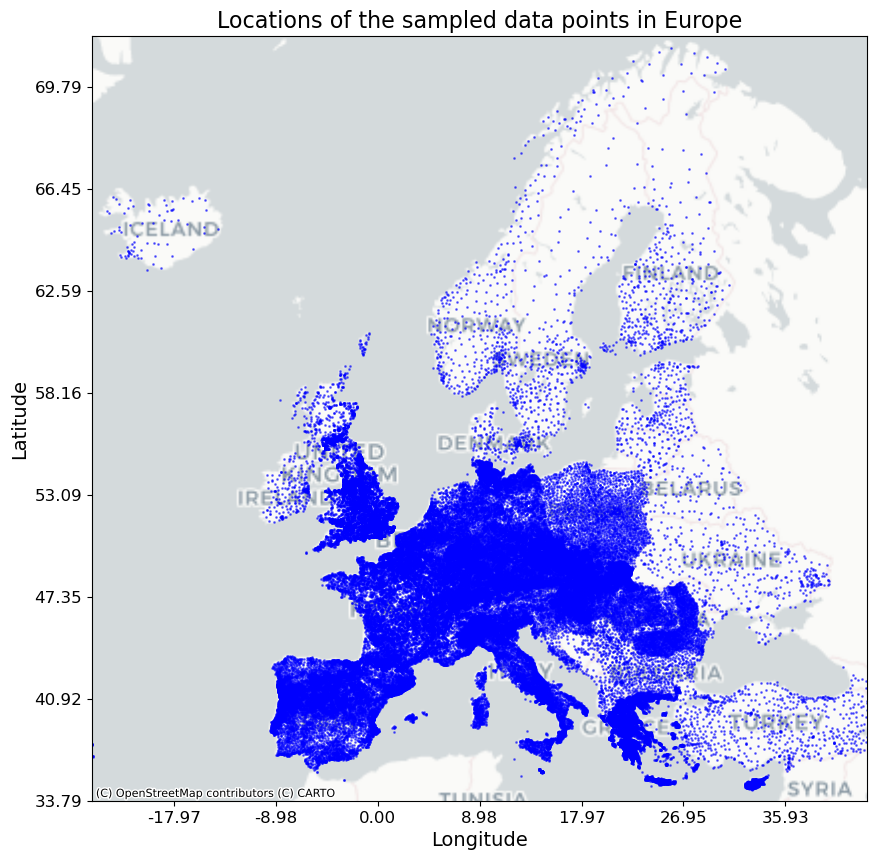}
    \caption{Our dataset consists of $95\,857$ sampled locations in European countries. The sampling of the locations is done based on the European Local Administrative Units framework.}
    \label{fig:europe_data}
\end{figure}

We construct a multi-view dataset of $95\,857$ unique locations distributed across the European continent. The sampling of locations is guided by population distribution, using the European Local Administrative Units (LAU) framework \citep{eurostat_lau}. \added{The location encoder learns spatial variation in more detail where the training data contains more sampled locations. Sampling proportionally to population therefore places this detail where most insured objects are located. The urban or rural character of a location is captured by the spatial views, so the sampling only determines how finely the location encoder can distinguish nearby locations, not what information the embeddings contain. For case studies that require finer detail in rural areas, such as agricultural insurance, the sampling scheme can be adapted to place more training locations in those regions.} We show the distribution of the dataset's locations in Europe in Figure \ref{fig:europe_data}.

\begin{figure}[ht!]
\centering
\begin{subfigure}[t]{0.31\linewidth}
    \centering
  \includegraphics[height=5cm, keepaspectratio]{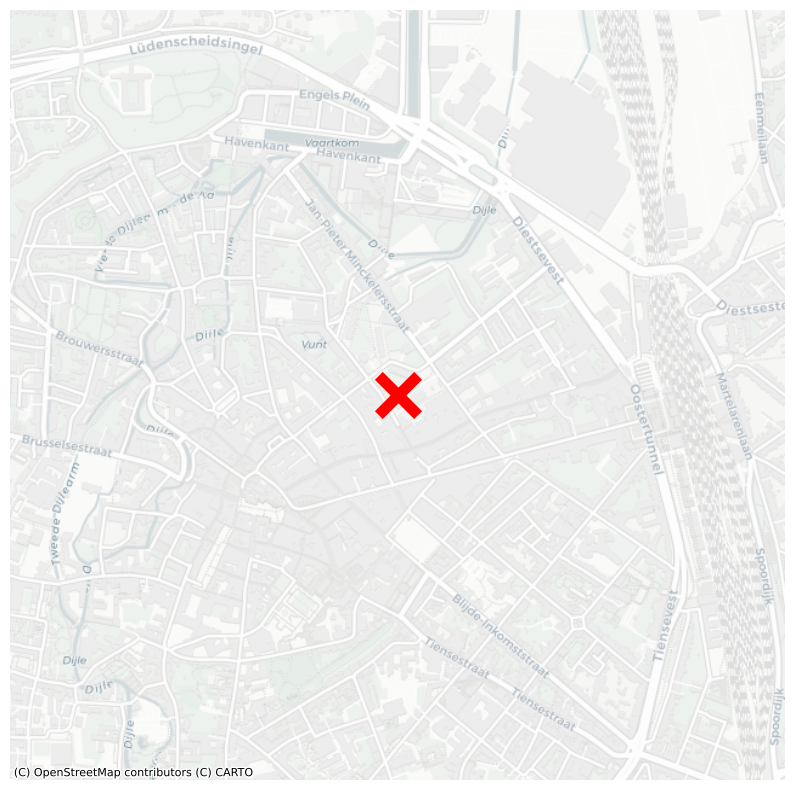}
  \caption{The coordinate pair $(\lambda,\phi) = (50.88173^{\circ}, 4.70591^{\circ})$ shown with a red cross. This corresponds to a location in Leuven, Belgium.}
  \label{fig:point_leuven}
\end{subfigure}%
\hspace{1em}
\begin{subfigure}[t]{0.31\linewidth}
    \centering
  \includegraphics[height=5cm, keepaspectratio]{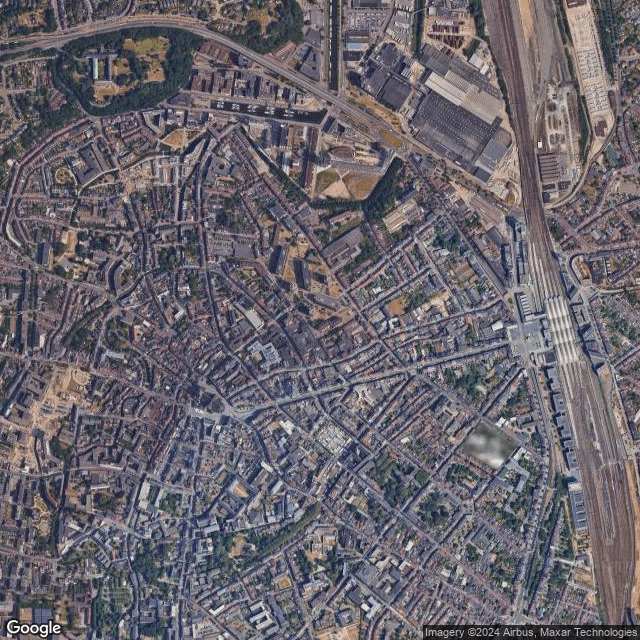}
  \caption{A Google Satellite image corresponding to the coordinate pair from panel (a). The data point $V^{\mathrm{GS}}$ represents the RGB pixel values from this image.}
  \label{fig:gs_leuven}
\end{subfigure}%
\hspace{1em}
\begin{subfigure}[t]{0.31\linewidth}
\centering
  \includegraphics[height=5cm, keepaspectratio]{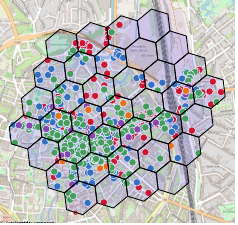}
  \caption{OpenStreetMap tags corresponding to the coordinate pair from panel (a). The data point $V^\mathrm{OSM}$ contains the count for each of the six categories in each of the hexagons.}
  \label{fig:osm_leuven}
\end{subfigure}
\caption{Illustration of one data point in the dataset $\mathcal{D}$, with coordinates $(\lambda,\phi) = (50.88173^{\circ}, 4.70591^{\circ})$ in Leuven, Belgium. Panel~(a) shows the location of the coordinate pair. Panel~(b) displays the associated Google Satellite image, where the matrix $V^{\mathrm{GS}}\in\mathbb{R}^{256\times256\times3}$ stores the associated RGB pixel values. Panel~(c) displays the corresponding OpenStreetMap features, where the matrix $V^{\mathrm{OSM}}\in\mathbb{N}^{37\times6}$ contains the counts of six categories of tags for each of the $37$ surrounding hexagons.}
\label{fig:gs_osm_example}
\end{figure}

Each location is defined by a latitude–longitude coordinate pair and enriched with two data views: a Google satellite image (GS) on the one hand and structured semantic context from OpenStreetMap (OSM) \citep{openstreetmap} on the other hand. Figure~\ref{fig:gs_osm_example} shows an example of the two spatial views for a location in Leuven, Belgium. We gather the satellite image using the Google Maps Static API \citep{google_maps_static_api}, as shown in Figure \ref{fig:gs_leuven}. The satellite image is centred on the coordinates of the location and has a resolution of $256\times256$ pixels in RGB format, resulting in a dimension of $256\times256\times3$. We gather OSM tags representing semantic information typically not visible in satellite imagery. Hereto, we define six categories of tags: \emph{shops}, \emph{tourism}, \emph{food \& drink}, \emph{health}, \emph{education} and \emph{public safety}. Appendix \ref{app:osmtags} contains all OpenStreetMap tags corresponding to each category. We aggregate occurrences of these tags in three rings of H3 hexagonal cells \citep{uber_h3} centred at each location. Using concentric rings around the centre, we capture both close-by and more distant amenities in a consistent spatial structure. This is illustrated in Figure~\ref{fig:osm_leuven}, which shows three rings of hexagons around the centre location with the occurrences of each category of tag in coloured points. Three rings of hexagons around a central location lead to $37$ hexagons in total, where the frequency of each of the six tag categories is counted for each hexagon.

Formally, let $P = 95\,857$ denote the total number of locations in our dataset. We define the dataset $\mathcal{D} = \{ (\lambda_p, \phi_p), V^{\mathrm{GS}}_p, V^{\mathrm{OSM}}_p \}_{p=1}^{P}$, where the pair $(\lambda_p, \phi_p)$ represents the geographical coordinates latitude and longitude of location $p$. Each location is associated with a satellite image represented by $V^{\mathrm{GS}}_p \in \mathbb{R}^{256 \times 256 \times 3}$, and $V^{\mathrm{OSM}}_p \in \mathbb{N}^{37 \times 6}$ represents the aggregated OpenStreetMap tag counts. Note that for a location $p$, the spatial views $V^{\mathrm{GS}}_p$ and $V^{\mathrm{OSM}}_p$ always \added{correspond to} the coordinates $(\lambda_p, \phi_p)$.

\section{Spatial embedding model architecture} \label{sec:model_archi}

Using the two spatial views introduced in Section \ref{sec:spatialdata}, we construct our spatial embedding model. Firstly, we introduce the two spatial view encoders and the structure of the multi-view fusion encoder as sketched in Figure \ref{fig:multiviewmodel}. Secondly, we detail the architecture of the location encoder. Thirdly, we provide an overview of the model's complete architecture, training, loss function, and examine the hyperparameter choices for training the spatial embedding model.

\subsection{Spatial view encoders and multi-view fusion} \label{sec:spatialencoders}

\begin{figure}[ht!]
    \centering
    \includegraphics[width=\linewidth]{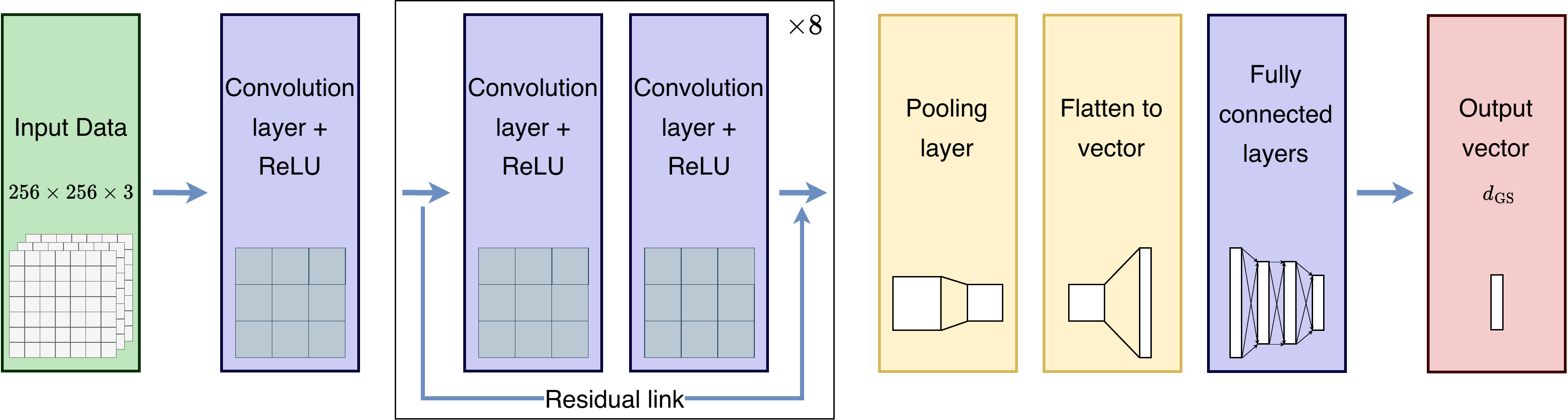}
    \caption{Illustration of the encoder used for the Google satellite imagery spatial view. We use a ResNet18 architecture \citep{He2016-ab}, pretrained following \citet{Wang2023}. The input image (green block) of dimension $256\times256\times3$ is processed by convolutional layers (\added{purple} blocks) that extract spatial patterns. These patterns are then pooled and flattened (yellow blocks) before passing through a fully connected network (rightmost \added{purple} block). The final embedding is a vector of dimension $d_{\mathrm{GS}}$, as shown in the red block. \added{Purple} blocks contain the trainable parameters of the model.}
    \label{fig:gs_flowchart}
\end{figure}

\paragraph{Google satellite encoder (GS)} The image encoder, denoted as $f^{\mathrm{GS}}(V^{\mathrm{GS}})$, maps a satellite image $V^{\mathrm{GS}} \in \mathbb{R}^{256\times256\times3}$ to a $d_{\mathrm{GS}}$-dimensional vector. As encoder $f^{\mathrm{GS}}$, we use a convolutional neural network, since such architectures are specifically designed to extract and combine patterns from image data. In particular, we adopt a ResNet18 model \citep{He2016-ab}, a convolutional neural network, trained on satellite imagery using momentum contrast \citep{He2019-so}. Momentum contrast is a self-supervised method that trains a network to represent similar inputs by similar vectors. \added{The network is pretrained on SSL4EO-S12 \citep{Wang2023}, a large-scale collection of satellite imagery assembled for self-supervised learning. \citet{Wang2023} report strong performance of the pretrained network on downstream remote sensing tasks. Starting from a pretrained network avoids training an image encoder from scratch, which would require far more data and computation. Among the published pretrained encoders, ResNet18 is the smallest architecture, which matches our aim of a framework that can be retrained on consumer-grade hardware.} The ResNet18 architecture consists of 17 convolutional layers with residual links to facilitate training \citep{He2016-ab}. \added{Each convolutional layer consists of several convolutional filters, meant to extract patterns from the input data. The number and size of these filters are defined by the ResNet18 architecture \citep{Wang2023} and are not tuned since we adopt the pretrained network directly.} The last convolutional layer is connected to a pooling layer and flattened into a vector, which is then passed through a fully connected network to produce the final embedding. Figure \ref{fig:gs_flowchart} illustrates this architecture. The input satellite image $V^{\mathrm{GS}}$ is represented by a green input block. It is processed through a sequence of convolutional layers (\added{purple} blocks), followed by pooling and flattening operations (yellow blocks). The resulting representation is then passed through a fully connected network (rightmost \added{purple} block), which outputs the final embedding vector of dimension $d_{\mathrm{GS}}$ shown in red.

We initialise the model with pretrained weights from \citet{Wang2023} and adjust the output dimension to $d_{\mathrm{GS}}$, a hyperparameter. The trainable parameters of $f^{\mathrm{GS}}$ lie in the convolutional filters and in the fully connected output network, represented as \added{purple} blocks in Figure \ref{fig:gs_flowchart}.

\paragraph{OpenStreetMap tags encoder (OSM)} We denote the encoder for the OpenStreetMap tags view as $f^{\mathrm{OSM}}(V^{\mathrm{OSM}})$ with the encoder mapping a hexagonal data input $V^{\mathrm{OSM}} \in \mathbb{N}^{37\times6}$ to a $d_{\mathrm{OSM}}$-dimensional vector. Similar to the encoder $f^{\mathrm{GS}}$, we use a convolutional neural network to extract patterns from the input data and represent those patterns in a vector. However, our input is defined on a hexagonal grid, whereas standard convolutional neural networks are designed for square-grid data such as images. Therefore, we adopt the mapping proposed by \citet{moussaid2023convolutional} for convolutional neural networks on hexagonal input data while preserving neighbourhood relations between cells. The mapping follows the h2 coordinate system \citep{snyder_hexcoord}, which minimises distortion by maintaining adjacency and approximate distance between neighbouring hexagons. Although alternative mappings are possible, this approach provides a consistent alignment between hexagonal neighbourhoods and square convolutional filters, allowing the use of standard convolutional layers without altering the underlying spatial structure of the data. \added{Alternative strategies exist to handle hexagonally structured inputs. The hexagonal cells can be shifted into a rectangular array combined with adapted convolution operations, as implemented in HexagDLy \citep{steppa2019hexagdly}. Convolutions that operate natively on the hexagonal lattice are possible as well \citep{hoogeboom2018hexaconv}. Both strategies require specialised convolution implementations.}

\begin{figure}[ht]
\centering
\begin{subfigure}[t]{0.48\linewidth}
    \centering
  \includegraphics[width=5cm, keepaspectratio]{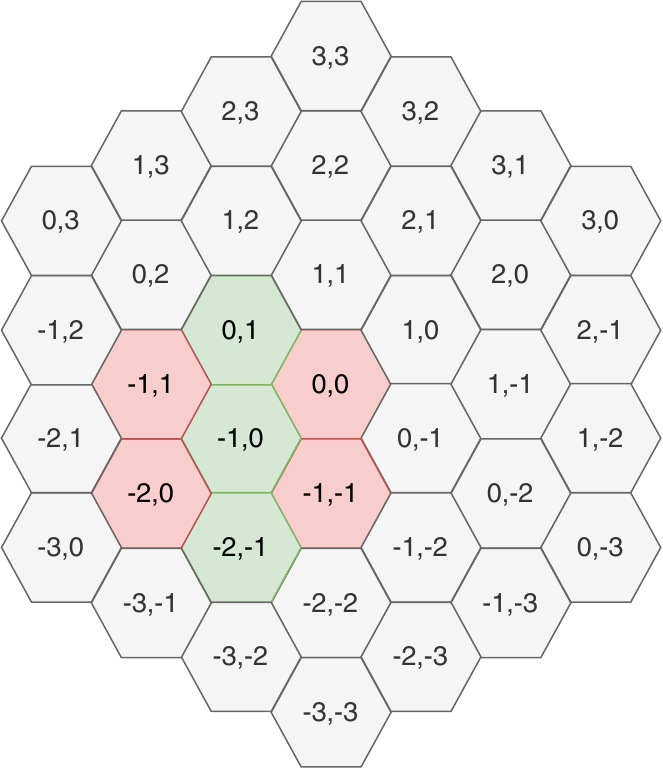}
  \caption{A hexagonal grid of $37$ cells (three rings). Coordinates are added for clarity. In colour, a convolutional filter of size $k=1$ (one ring) is shown.}
  \label{fig:hex_coord}
\end{subfigure}%
\hspace{1em}
\begin{subfigure}[t]{0.48\linewidth}
\centering
  \includegraphics[width=5cm, keepaspectratio]{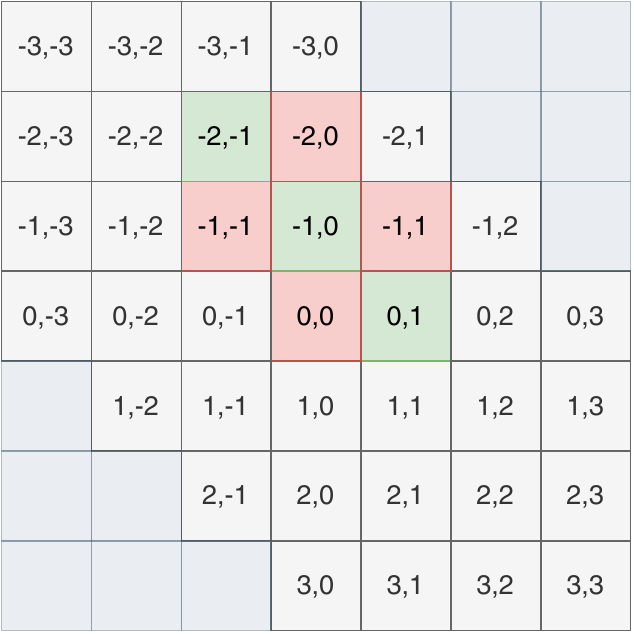}
  \caption{The corresponding $7\times7$ square grid obtained from the transformation of the $37$ hexagonal cells following \citet{moussaid2023convolutional}. Coordinates indicate how hexagonal cells map to the square representation. In colour, the $k=1$ filter is shown after transformation, corresponding to a $3\times3$ filter with the upper-right and lower-left corners masked.}
  \label{fig:hex_coord_square}
\end{subfigure}
\caption{Illustration of the transformation from hexagonal to square representation. A convolutional filter of size $k=1$ on the hexagonal grid (left) corresponds to a $3\times3$ filter on the transformed square grid (right), with the upper-right and lower-left corners masked.}
\label{fig:hex_convolutions}
\end{figure}

We illustrate this transformation with a grid of $37$ hexagons in Figure \ref{fig:hex_coord}, which is mapped to a $7\times7$ matrix in Figure \ref{fig:hex_coord_square} following the h2 coordinate system \citep{snyder_hexcoord}, which assigns two-axis coordinates to the cells of a hexagonal grid. The upper and lower corners of the matrix are masked, meaning that these entries are fixed to zero because no corresponding hexagons exist in those positions. For clarity, coordinates are added to each hexagonal cell and their correspondence to the square matrix is shown. A convolutional filter covering one ring of hexagons is illustrated with red and green cells in Figure \ref{fig:hex_coord}. In the transformed representation, this filter corresponds to a $3\times3$ convolutional filter, where the upper-right and lower-left corners are masked, shown with the red and green cells in Figure \ref{fig:hex_coord_square}.  Generally, for a convolutional filter of $k$-rings around a central hexagon, the corresponding standard convolutional filter has a size of $2k+1$, with zeros masking the upper-right and lower-left corners of size $k$. Hence, the encoder $f^{\mathrm{OSM}}$ is built using a standard convolutional architecture on transformed input data and by fixing the upper right and lower left corners of the convolutional filters to zero.

\begin{figure}[ht!]
    \centering
    \includegraphics[width=\linewidth]{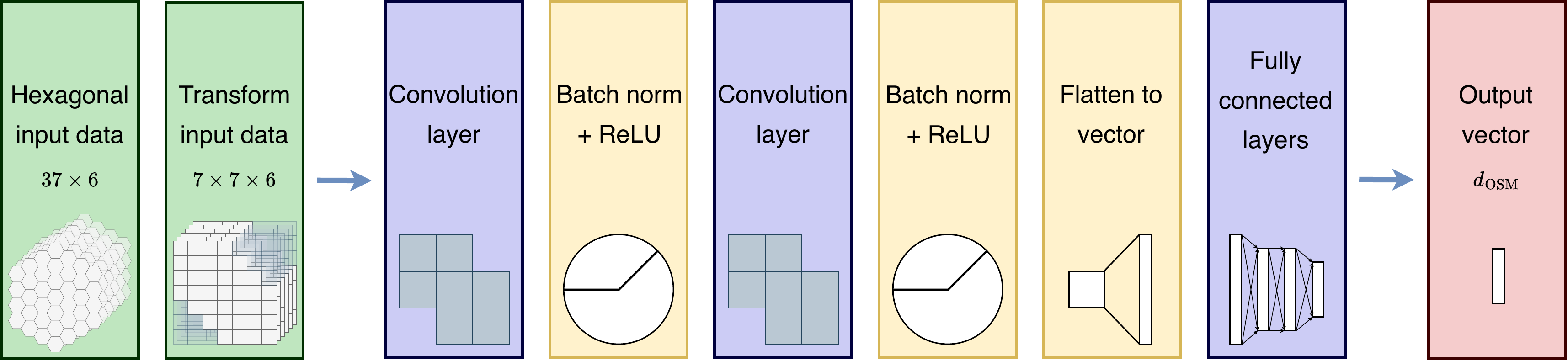}
    \caption{Illustration of the encoder used for the OpenStreetMap spatial view. The hexagonal input data is first transformed into a rectangular format. Next, there are two convolutional layers, each using masked convolutions. Finally, a feed-forward neural network outputs the encoded input. Each \added{purple} block in the flowchart contains trainable elements to be optimised in the training of the spatial encoder model.}
    \label{fig:osm_flowchart}
\end{figure}

The encoder $f^{\mathrm{OSM}}$ is illustrated in Figure \ref{fig:osm_flowchart}. The input is a hexagonal grid consisting of $37$ cells, where each cell counts the number of tags for each of the six OSM categories under consideration, resulting in an input dimension of $37\times6$. We apply the transformation on our hexagonal input, which results in a square matrix of dimension $7\times7\times6$. On the transformed input, we apply two convolutional layers. \added{Two layers suffice to cover the full spatial extent of the small $7\times7$ input while keeping the number of trainable parameters limited. The number of convolutional filters in each layer, together with the size of the filters, is treated as a tuning parameter, selected during the preliminary testing described in Appendix \ref{app:hypertesting} and guided by training stability and the representational quality of the resulting embeddings rather than formal optimisation.} Each convolutional layer is followed by normalisation and ReLU activation to facilitate training. The output of the second convolutional layer is then flattened into a vector and passed through a feed-forward neural network to produce the final representation of dimension $d_{\mathrm{OSM}}$. This architecture enables the encoder to learn from the spatial structure of the OSM data, resulting in an output vector that accurately represents the spatial patterns in the input data.

The encoder $f^{\mathrm{OSM}}$ has trainable parameters in the filters of each of the two convolutional layers and in the fully connected network that produces the output. Figure \ref{fig:osm_flowchart} shows these components as \added{purple} blocks. The output dimension $d_{\mathrm{OSM}}$ is a hyperparameter.

\paragraph{Multi-view fusion}

We combine the embedding vectors produced by the spatial view encoders $f^{\mathrm{GS}}$ and $f^{\mathrm{OSM}}$ into a single joint spatial representation by use of a multi-view fusion encoder $f^{\mathrm{MV}}$. The two embeddings are concatenated and passed through a feed-forward neural network with fully connected layers, which outputs a vector of dimension $d$. These layers contain trainable weights that are optimised jointly with the rest of the spatial embedding model. The depth of the network and the width of each layer are treated as hyperparameters in the model design.

\subsection{Location encoder} \label{sec:locencoder}

\begin{figure}[ht!]
    \centering
    \includegraphics[width=\linewidth]{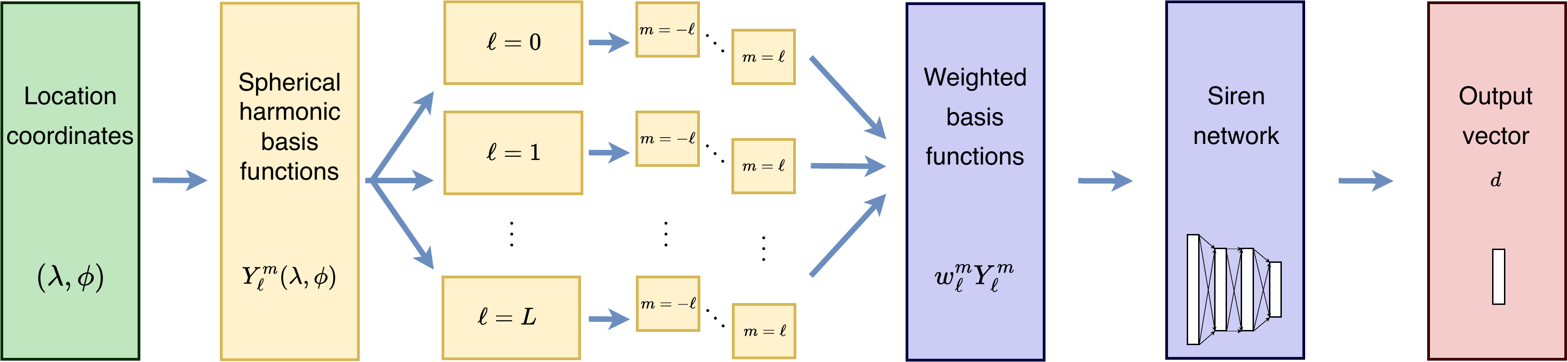}
    \caption{Illustration of the used location encoder. The weighted spherical harmonic basis functions $w_{\ell}^mY_{\ell}^m$ are inputs to a feed-forward neural network using sinusoidal activation functions, called the Siren net \citep{Ruswurm2023}. Each \added{purple} block in the flowchart contains trainable elements that can be optimised during the training of the spatial encoder model.}
    \label{fig:loc_flowchart}
\end{figure}

Each data point's coordinates $(\lambda,\phi)$ are embedded onto a $d$-dimensional vector by a location encoder, as illustrated in Figure \ref{fig:loc_flowchart}. We denote the location encoder as

\begin{equation}
    f^{\mathrm{LE}}:\mathcal{D}\to\mathbb{R}^d: (\lambda,\phi) \mapsto f^{\mathrm{LE}}\left(\lambda,\phi\right).
\end{equation}

We use the proposed location encoder by \citet{Ruswurm2023}, where $f^{\mathrm{LE}}(\lambda,\phi) = \mathrm{NN}(\mathrm{PE}(\lambda,\phi))$, consisting of a positional encoder $\mathrm{PE}(\lambda,\phi)$ and a neural network part $\mathrm{NN}(\cdot)$. \citet{Ruswurm2023} apply this setup to learn spatial patterns from weather variables, i.e., temperature and air pressure data, for use in a regression model, and from satellite imagery for land-use classification and land-ocean segmentation.

The positional encoder is defined as
\begin{equation}
    \mathrm{PE}(\lambda,\phi) = \left[w_{\ell}^m Y_{\ell}^m(\lambda,\phi)\right]_{\ell=0,\dots,L;\; m=-\ell,\dots,\ell},
    \label{eq:sphericalharmonics}
\end{equation}

where $w_{\ell}^m$ are weights and $Y_{\ell}^m$ are the orthogonal spherical harmonic basis functions associated with Legendre polynomials $P_{\ell}^m$ of degree ${\ell}$ and order $m$, see \citet{Ruswurm2023}. These spherical harmonic functions encode spatial locations on a spherical surface by decomposing them into a sum of orthogonal components. The orthogonality ensures that different spatial components are captured separately, allowing the model to represent complex spatial patterns at global and localised scales. The neural network part, known as the Siren network \citep{SitzmannVincent2020INRw, klemmer2023satclip, Ruswurm2023}, takes as input the weighted spherical harmonic functions $w_{\ell}^mY_{\ell}^m(\lambda,\phi)$. These are passed through feed-forward layers with sinusoidal activation functions, which allow the network to represent spatial patterns smoothly on a sphere.

Figure \ref{fig:loc_flowchart} sketches the architecture of the location encoder. The input is a coordinate pair $(\lambda, \phi)$, represented by the green block. We evaluate the spherical harmonic basis functions (yellow blocks), for $\ell=0,\ldots,L$ and for each $\ell$ over $m=-\ell,\ldots,\ell$, using this coordinate pair as input. The weighted spherical harmonic basis functions (first \added{purple} block) serve as inputs to the feed-forward neural network with sinusoidal activation functions, known as the Siren network (second \added{purple} block). The output of the Siren network is the location embedding of the coordinate pair. 

The weights of the spherical harmonic basis functions and the weights in the Siren network are trainable parameters of the location encoder. Increasing the number of basis functions $L$ allows the positional encoder to capture more fine-grained spatial detail, but also increases computational cost. To effectively utilise this additional detail, a larger value for $L$ typically implies a larger value for $d$, ensuring the encoder has sufficient capacity to represent the information \citep{klemmer2023satclip}. The values of $L$, $d$, and the architecture of the Siren network are treated as hyperparameters.

\subsection{Putting it all together: model architecture, training, and applications}

\begin{figure}[ht!]
    \centering
    \includegraphics[width=\linewidth]{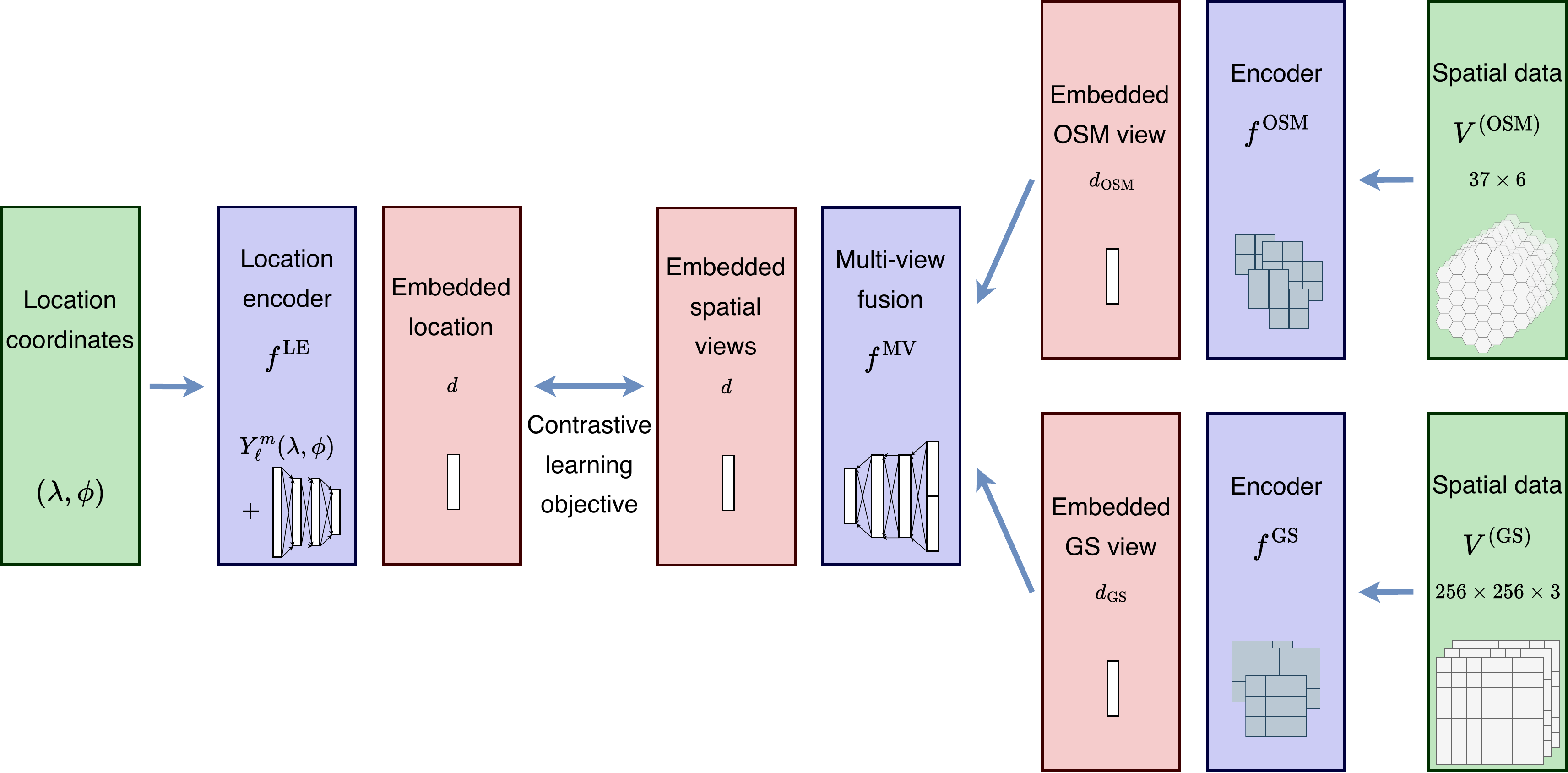}
    \caption{Architecture of the multi-view spatial embedding model with two spatial views as inputs. The left side shows the location encoder $f^{\mathrm{LE}}$ (Figure \ref{fig:loc_flowchart}), which embeds the coordinates $(\lambda,\phi)$. On the right, the Google satellite imagery and OpenStreetMap tags are processed by their respective encoders $f^{\mathrm{GS}}$ and $f^{\mathrm{OSM}}$ (Figures \ref{fig:gs_flowchart} and \ref{fig:osm_flowchart}). Their outputs are combined in the multi-view fusion encoder $f^{\mathrm{MV}}$. The resulting spatial view embedding is aligned with the location embedding using contrastive learning. \added{Purple} blocks indicate trainable components that are optimised during model training.}
    \label{fig:mvenc_flowchart}
\end{figure}

Figure \ref{fig:mvenc_flowchart} illustrates the complete architecture for our spatial embedding model using two spatial views as inputs. The left-hand side represents the encoding of the location coordinates $(\lambda, \phi)$ by the location encoder $f^{\mathrm{LE}}$ as discussed in Section \ref{sec:locencoder}, resulting in an embedded location vector of dimension $d$. On the right, the spatial view encoders, as discussed in Section \ref{sec:spatialencoders}, process the two input data sources: the OpenStreetMap tags, encoded by $f^{\mathrm{OSM}}(V^{\mathrm{OSM}})$, into a vector of dimension $d_{\mathrm{OSM}}$, and the Google satellite imagery, encoded by $f^{\mathrm{GS}}(V^{\mathrm{GS}})$, into a vector of dimension $d_{\mathrm{GS}}$. These two spatial view embeddings are then combined by the multi-view fusion encoder $f^{\mathrm{MV}}\left(f^{\mathrm{GS}}(V^{\mathrm{GS}}), f^{\mathrm{OSM}}(V^{\mathrm{OSM}})\right)$ into a single joint embedding of dimension $d$, with typically $d \le d_{\mathrm{GS}} + d_{\mathrm{OSM}}$.

\sloppy From the dataset $\mathcal{D}$, introduced in Section \ref{sec:spatialdata}, we construct a training dataset $\mathcal{D}^T=\{(\lambda_p,\phi_p),V^\mathrm{GS}_p,V^\mathrm{OSM}_p\}_{p=1}^{P^T}$ of size $P^T$, containing $90\%$ of the dataset $\mathcal{D}$ sampled randomly. The remaining $10\%$ of data forms the validation set $\mathcal{D}^V$ and is used for early stopping in the training of the model. The training objective is to align the embedded locations with the embedded spatial views. For each $p = 1, \ldots, P^T$, we denote the embedding vectors as $z$ for notational clarity:
\[
z^\mathrm{LE}_p = f^\mathrm{LE}(\lambda_p, \phi_p), \quad 
z^\mathrm{MV}_p = f^\mathrm{MV}\left(f^\mathrm{GS}(V^\mathrm{GS}_p), f^\mathrm{OSM}(V^\mathrm{OSM}_p)\right).
\]
The contrastive loss function $\mathscr{L}$ is constructed to maximise the similarity between the so called positive pairs $z^\mathrm{LE}_p$ and $z^\mathrm{MV}_p$, and to minimise the similarity across all mismatched pairs $z^\mathrm{LE}_p$ and $z^\mathrm{MV}_{p'}$ for $p' \neq p$, see \citet{Radford2021-et}. The contrastive loss over all elements in the training dataset is defined as
\[
\mathscr{L} = \frac{1}{2P^T} \sum_{p=1}^{P^T} \left[
    -\log \frac{\exp\left( \text{sim}(z^\mathrm{LE}_p, z^\mathrm{MV}_p)/\tau \right)}
    {\sum_{p'=1}^{P^T} \exp\left( \text{sim}(z^\mathrm{LE}_p, z^\mathrm{MV}_{p'})/\tau \right)}
    -\log \frac{\exp\left( \text{sim}(z^\mathrm{MV}_p, z^\mathrm{LE}_p)/\tau \right)}
    {\sum_{p'=1}^{P^T} \exp\left( \text{sim}(z^\mathrm{MV}_p, z^\mathrm{LE}_{p'})/\tau \right)}
\right],
\]
where $\text{sim}(a,b) = \frac{a \cdot b}{\lVert a\rVert\,\lVert b\rVert}$ denotes the normalised dot-product and $\tau$ is a so-called temperature hyperparameter. The two similarity terms ensure that both the alignment from location to spatial embeddings, $\text{sim}(z^\mathrm{LE}_p, z^\mathrm{MV}_p)$, and from spatial to location embeddings, $\text{sim}(z^\mathrm{MV}_p, z^\mathrm{LE}_p)$, is maximised. Training the complete architecture of our spatially embedded model consists of finding the optimal values for the trainable parameters so that the contrastive loss $\mathscr{L}$ on the training dataset $\mathcal{D}^T$ is minimised. The trainable parameters in the model architecture are illustrated with \added{purple} blocks in Figure \ref{fig:mvenc_flowchart}. Minimising the contrastive loss encourages the location embeddings $z^\mathrm{LE}_p$ and the multi-view embeddings $z^\mathrm{MV}_p$ of the same point $p$ to be similar, while embeddings of different locations remain distinct. In this way, the positional patterns represented by the spherical harmonics are aligned with the information extracted from the spatial views. The resulting embedded locations thus encode both the geographic position and the spatial content, allowing the model to represent locations in a manner that reflects where they are as well as what is observed there.

\begin{figure}[ht!]
\centering
\includegraphics[width=\linewidth]{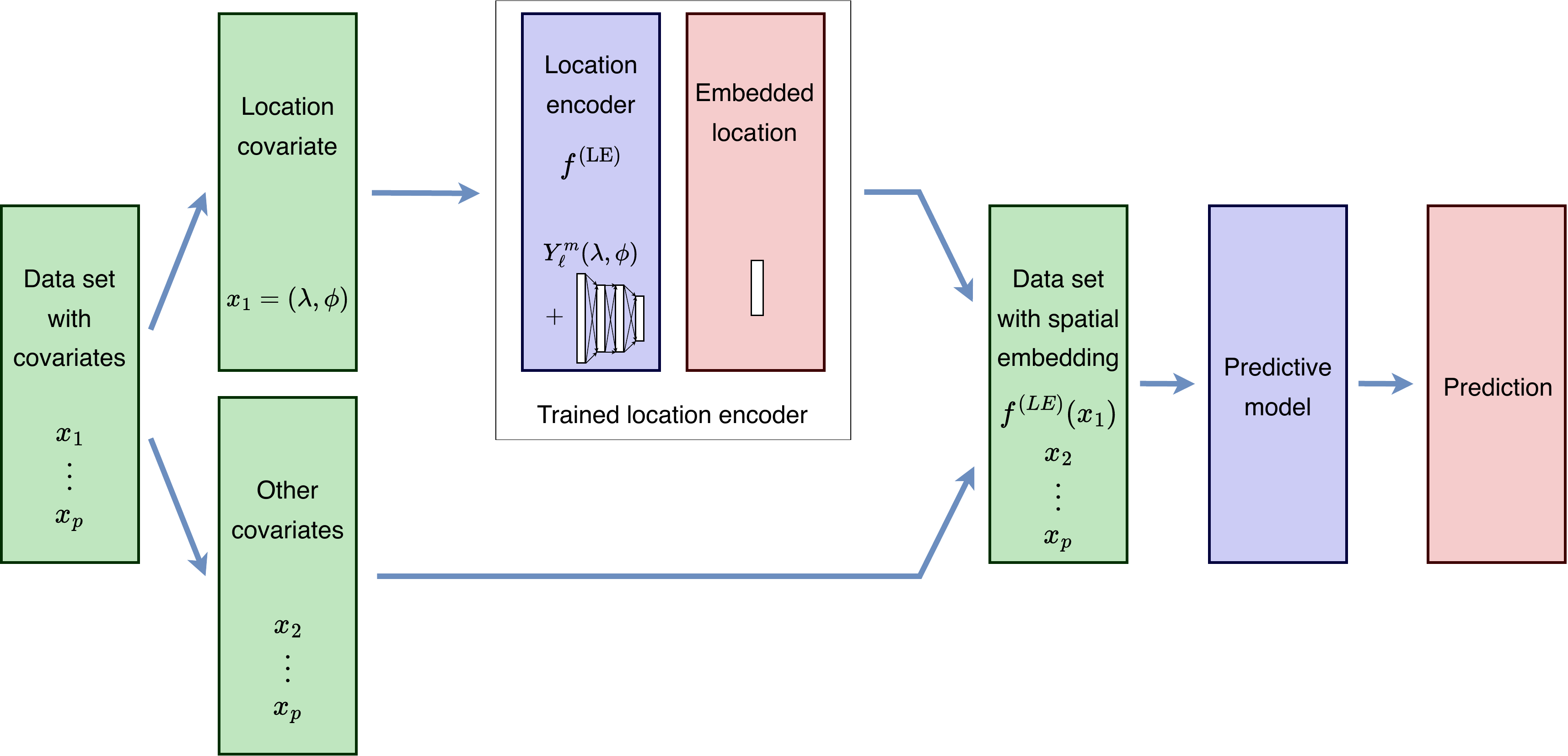}
\caption{Application of the trained spatial embedding model in a case study where geographical coordinates are used as a covariate in a predictive model. By replacing the raw coordinate $x_1$ with its spatial embedding $f^{\mathrm{LE}}(x_1)$, the model incorporates not only geographic position but also the spatial information observed at that location through the embedding.}
\label{fig:application_flowchart}
\end{figure}

In a case study involving geographical locations, the trained spatial embedding model can be applied to represent each location by its embedding. For a dataset consisting of covariates $x_1,\ldots,x_p$, where $x_1$ represents geographical coordinates, the covariate $x_1$ can be replaced with its embedding $f^{\mathrm{LE}}(x_1)$, as illustrated in Figure \ref{fig:application_flowchart}. These embeddings act as compact descriptors that integrate both geographic position and the spatial information captured from multiple data views, and can be used as input features in downstream analyses such as regression models.

\subsection{Hyperparameter choices and trained model architectures} \label{sec:hyperparam}

In the construction and training of our spatial embedding model architecture, several hyperparameters have to be specified. The OpenStreetMap encoder $f^\mathrm{OSM}$ consists of two convolutional layers, the first with $8$ filters and the second with $16$ filters, both with kernel size $k=1$. \added{These values were selected during preliminary testing, as detailed in Appendix \ref{app:hypertesting}.} The multi-view fusion encoder is implemented as a single dense layer consisting of $128$ neurons. The location encoder uses $L=64$ spherical harmonic basis functions in the positional encoder, followed by a two-layer feed-forward network with $128$ neurons each. Optimisation is performed with the Adam algorithm \citep{Kingma2015-tb} using a learning rate of $10^{-5}$ and weight decay of $0.01$. Training is carried out for $500$ epochs with a batch size of $128$. The contrastive loss employs a temperature parameter of $\tau = 0.07$. The hyperparameter choices either follow the values proposed in \citet{klemmer2023satclip} or were determined during preliminary testing\added{, guided by training stability and the representational quality of the resulting embeddings rather than formal optimisation,} in the model construction phase. More details can be found in Appendix \ref{app:hypertesting}.

For the experimental study in Section \ref{sec:frpricing}, we study the effect of the overall embedding dimension $d$, and the intermediate dimensions $d_{\text{GS}}$ and $d_{\text{OSM}}$ of the Google Satellite and OpenStreetMap encoders on the resulting spatial embeddings. To this end, we construct five different embedding models, each corresponding to a different combination of $d$, $d_{\text{GS}}$, and $d_{\text{OSM}}$. Table \ref{tab:hyperparam} shows the values of $d$, $d_{\text{GS}}$, and $d_{\text{OSM}}$ for each of the five embedding models, which are compared in the experimental study in Section \ref{sec:frpricing}.

\setlength{\extrarowheight}{3pt} 
\begin{table}[ht]
\centering
\begin{NiceTabular}{l>{\raggedleft\arraybackslash}m{2.1cm}>{\raggedleft\arraybackslash}m{1.9cm}>{\raggedleft\arraybackslash}m{1.9cm}}[
code-before = \rowcolor[HTML]{FFFFFF}{1,2,4,6}
              \rowcolor[HTML]{FFFFFF}{3,5}
]
\toprule
Model name & Embedding dimension $d$ & GS spatial view & OSM spatial view \\
\noalign{\hrule height 0.3pt}
\texttt{EU8\_GS32\_OSM32}   & $8$  & $d_{\text{GS}} = 32$ & $d_{\text{OSM}} = 32$  \\ 
\texttt{EU16\_GS32\_OSM16}  & $16$ & $d_{\text{GS}} = 32$ & $d_{\text{OSM}} = 16$ \\ 
\texttt{EU16\_OSM16}        & $16$ & $d_{\text{GS}} = 0$  & $d_{\text{OSM}} = 16$ \\  
\texttt{EU32\_GS96\_OSM32}  & $32$ & $d_{\text{GS}} = 96$ & $d_{\text{OSM}} = 32$ \\  
\texttt{EU64\_GS64}         & $64$ & $d_{\text{GS}} = 64$ & $d_{\text{OSM}} = 0$  \\
\bottomrule
\end{NiceTabular}
\caption{Overview of the five spatial embedding models used in the experimental study. Each model varies the overall embedding dimension $d$ and the intermediate dimensions $d_{\text{GS}}$ and $d_{\text{OSM}}$, while all other hyperparameters are fixed. Model acronyms are constructed as follows: the prefix \texttt{EU} indicates that the model is trained on the European dataset introduced in Section~\ref{sec:spatialdata}; this is followed by the overall embedding dimension $d$; then \texttt{GS} with the dimension of the Google Satellite encoder $d_{\text{GS}}$; and finally \texttt{OSM} with the dimension of the OpenStreetMap encoder $d_{\text{OSM}}$.}
\label{tab:hyperparam}
\end{table}

\added{The framework can be used in two ways. A first option is to apply the location encoder of one of the five trained models in Table \ref{tab:hyperparam}, published on Hugging Face \citep{holvoet_mvmodels_2025}, to any dataset with coordinates in Europe. A second option is to retrain the framework with spatial views tailored to a specific application, such as building characteristics from a commercial data provider or portfolio-specific exposure information.}

All training is conducted using Python on a desktop computer with a single NVIDIA RTX 4070 GPU (12 GB), an Intel i9-13900K CPU, and 128GB of RAM. \added{The implementation uses PyTorch and PyTorch Lightning for the construction and training of the embedding model, torchgeo and rasterio for the satellite imagery, and pandas and NumPy for the data handling. The GitHub repository at \url{https://github.com/freekholvoet/MultiviewSpatialEmbeddings} lists the complete set of dependencies with version numbers.} Training the model on the full European dataset required approximately 12 hours, depending on hyperparameters. Calculating the embeddings for a new coordinate, as illustrated in Figure \ref{fig:application_flowchart}, takes under 1 millisecond.

\section{Experimental study}\label{sec:frpricing}

\leavevmode\added{We demonstrate the practical value of the spatial embeddings in two case studies. The first, in Section \ref{sec:housecase}, predicts real estate prices in France. We use this application for two reasons. First, the dataset is publicly available and contains exact coordinates per observation, which supports the reproducibility of our results, whereas insurance claims data at this level of spatial granularity are typically confidential. Second, the relationship between location and transaction prices is strong and well documented, which allows us to verify the spatial effects captured by the embeddings. The second case study, in Section \ref{sec:flood_casestudy}, models flood claim counts across Belgian postal codes.}

\subsection{Case study on real estate pricing}\label{sec:housecase}

\subsubsection{Real estate transactions dataset}\label{sec:fr_eda}

We use a real estate transaction dataset published by \citet{etalab_dvf}, called the DVF data base. Our dataset contains $200\,000$ property sales in France between $2017$ and $2023$. Each record in the dataset describes one transaction with the sale price and descriptive variables about the property. Table~\ref{tab:case1_vardescription} provides an overview of the variables in the dataset, including their descriptions and whether they are continuous or categorical.

\begin{table}[ht!]
    \centering
    \begin{NiceTabular}[t]{m{2.8cm}m{4.8cm}m{7.6cm}}[
            code-before = \rowcolor[HTML]{FFFFFF}{1,2,4,6,8}
            \rowcolor[HTML]{FFFFFF}{3,5,7,9}
        ]
        \toprule
        \textbf{Variable} & \textbf{Original name (Etalab)} & \textbf{Description}\\
        \noalign{\hrule height 0.3pt}
        \texttt{price} & \texttt{valeur\_fonciere} & Transaction price at which the property was sold, as a continuous variable. \\ 
        \texttt{type} & \texttt{type\_local} & Type of property, as a categorical variable with three levels: apartment (type 1), industrial/commercial (type 2), or house (type 3). \\ 
        \texttt{land\_use} & \texttt{code\_nature\_culture} & Land-use classification code, as a categorical variable with 27 levels. See Appendix~\ref{app:landuse} for the codes and their meaning. \\  
        \texttt{rooms} & \texttt{nombre\_pieces\_principale} & Number of rooms in the property, as a continuous variable. \\ 
        \texttt{building\_area} & \texttt{surface\_reelle\_bati} & Surface area of the building in $m^2$, as a continuous variable. \\ 
        \texttt{land\_area} & \texttt{surface\_terrain} & Surface area of the land in $m^2$, as a continuous variable. \\ 
        \texttt{date} & \texttt{date\_mutation} & Date of transaction, represented as three continuous variables: year, month, and day. \\ 
        \texttt{coordinates} & \texttt{latitude}, \texttt{longitude} & Coordinates of the property, expressed as two continuous variables. \\ 
        \bottomrule
    \end{NiceTabular}
    \caption{Description of the variables in the French DVF real estate transaction dataset. The column \emph{Variable} lists the variable names used in this paper, while the column \emph{Original name (Etalab)} shows the dataset field names as defined in \citet{etalab_dvf}. Each description specifies whether the variable is continuous or categorical. For categorical variables the levels are listed.}
    \label{tab:case1_vardescription}
\end{table}

\begin{figure}[ht!]
    \centering
    \includegraphics[width=\linewidth]{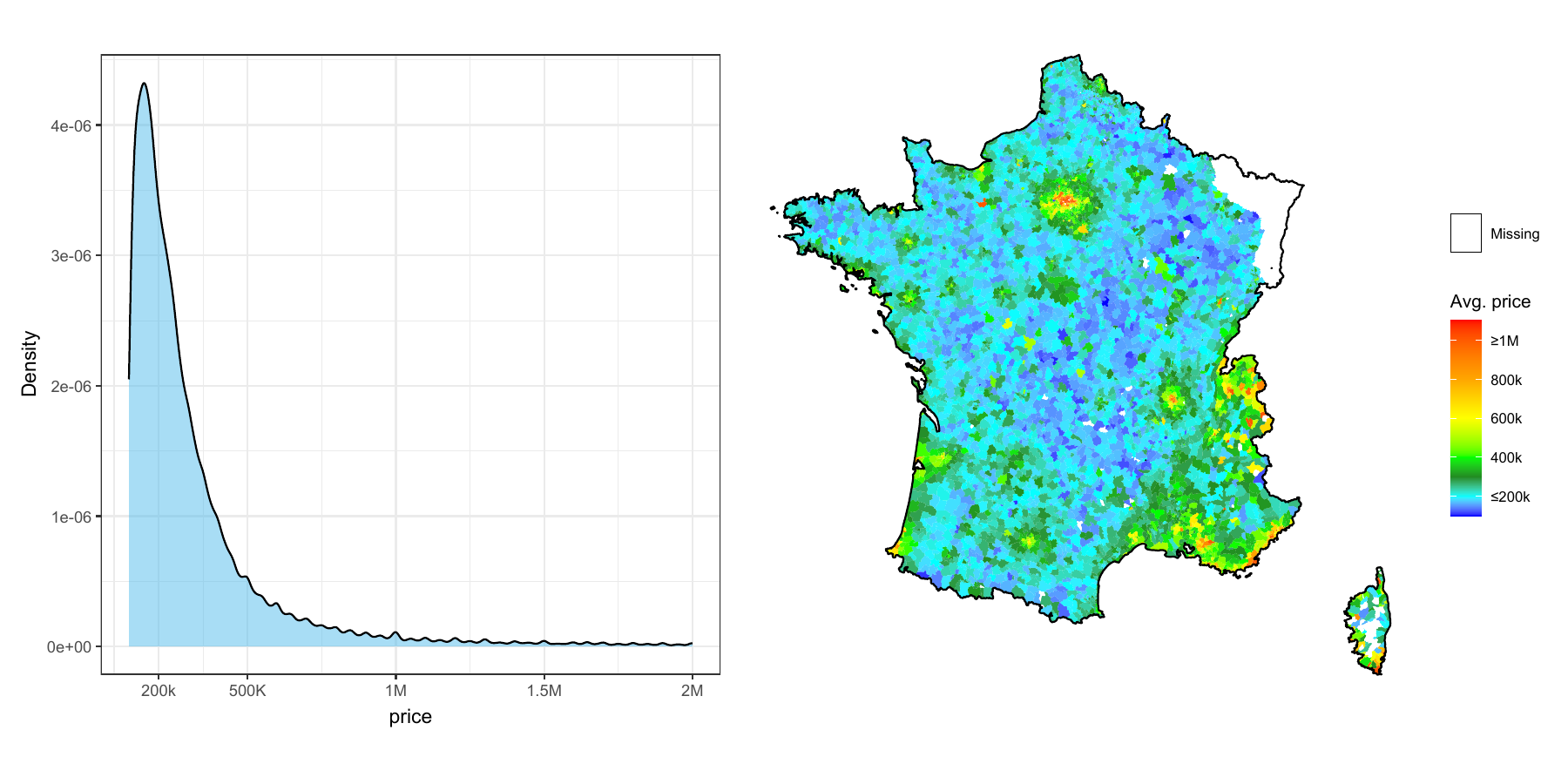}
    \caption{Plot of the property values in the dataset. The left-hand side shows a density plot of the property values. The right-hand side shows the average property value by postal code.}
    \label{fig:case1_responsevar}
\end{figure}

The case study constructs a regression model using the variables in Table \ref{tab:case1_vardescription} as covariates with \texttt{price} as the response variable. The left-hand side of Figure \ref{fig:case1_responsevar} shows the distribution of the response variable in the dataset, and the right-hand side shows the average property value by postal code. All transaction prices lie between \EUR{100\,000} and \EUR{2\,000\,000}. The average transaction value of properties is \EUR{310\,190}, with $90\%$ of properties having a transaction value between \EUR{112\,000} and \EUR{800\,000}. The areas with a higher average transaction value include the Paris district, the south-east and south-west coast areas, and the French Alps bordering Switzerland. The centre of France features large areas with lower average transaction prices in the regions of Limousin, Auvergne, and Bourgogne, which are characterised by extensive areas of nature reserves and agricultural land. We observe a higher average transaction price in cities such as Orleans, Lyon, and Toulouse. The dataset does not contain properties in the departments Bas-Rhin, Haut-Rhin, and Moselle.

\added{The transactions in the dataset date from $2017$ to $2023$, whereas the satellite imagery and OpenStreetMap tags used to train the embedding models were collected at the time of this study. The embeddings therefore describe the current spatial context of each location rather than the context at the moment of sale. This temporal misalignment is a limitation of this specific demonstration and not of the framework. In practice, the embedding model can be retrained per calendar year, or per any other relevant period, with temporally matched data. Historical satellite imagery is available through Google Earth Engine \citep{gorelick2017google}, and past states of OpenStreetMap can be reconstructed from edit timestamps.}

\begin{figure}[ht!]
    \centering
    \includegraphics[width=\linewidth]{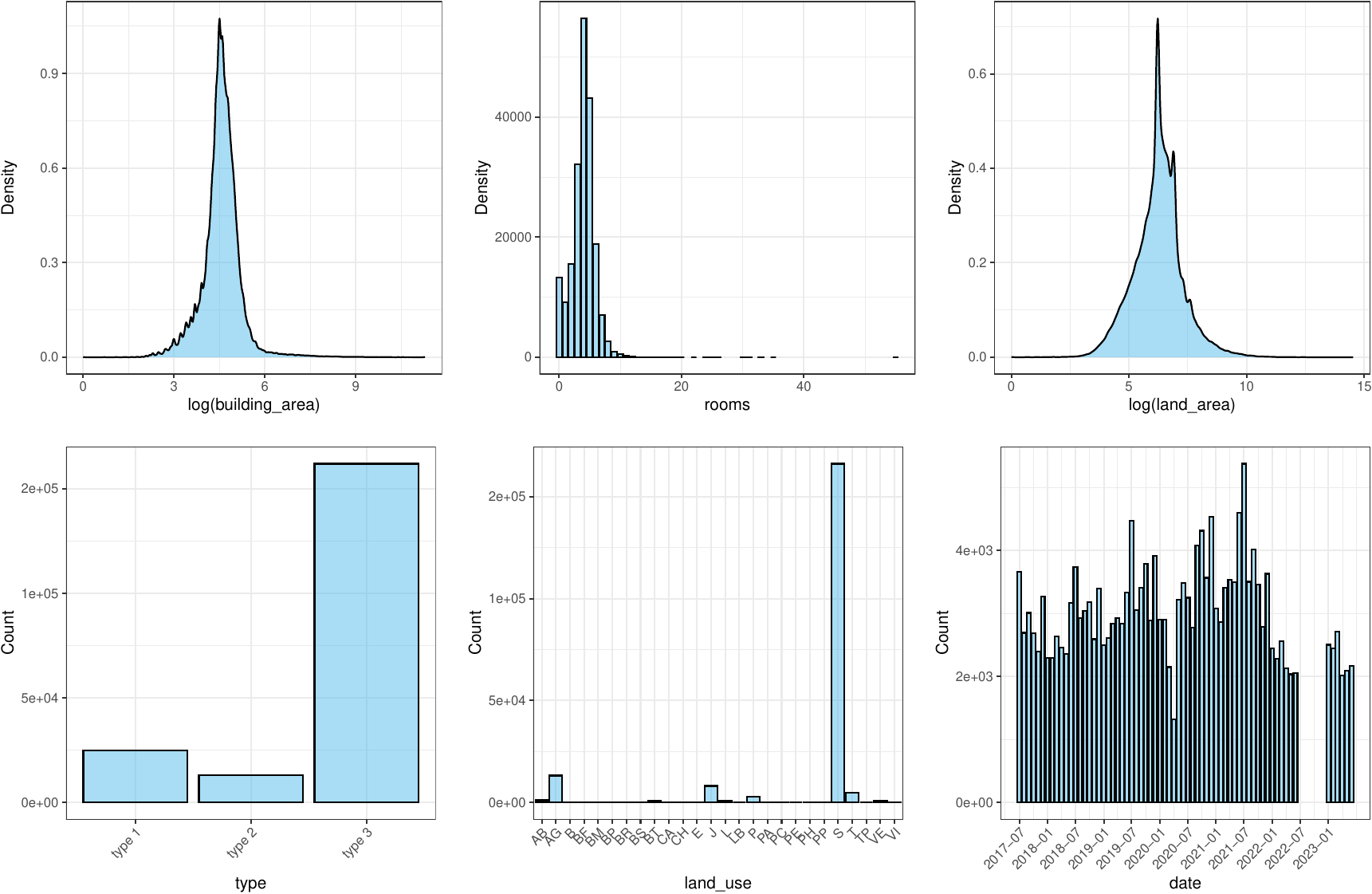}
    \caption{Exploratory data analysis of the French real-estate transaction dataset. From left to right, top row: density of $\log(\texttt{building\_area})$, bar plot of $\texttt{rooms}$, and the density of $\log(\texttt{land\_area})$. Bottom row: bar plot of $\texttt{type}$, bar plot of $\texttt{land\_use}$, and a bar plot showing the number of transactions per month over time.}
    \label{fig:case1_allvariables}
\end{figure}

Figure \ref{fig:case1_allvariables} shows the distribution of the other variables in the French real estate dataset. The properties in the dataset are on average \SI{135}{\metre\squared}, with $95\%$ of properties being below \SI{220}{\metre\squared}. The properties have on average four rooms, and a land area of \SI{1012}{\metre\squared}. Around $81\%$ of properties are classified as houses (Type 3), $12.5\%$ are apartments (Type 1), and the others are industrial or commercial properties (Type 2). Most land use classifications are building lots (class S). More properties were sold in $2021$ in comparison to $2020$, and no sales were recorded in the latter half of $2022$.

We denote the French real estate transaction dataset as $\mathcal{D}^{\mathrm{FR}} = \left(\boldsymbol{x}_i,y_i\right)_{i=1}^{200\,000}$. Each index $i$ represents one property in the dataset, with the vector $(x_{i,1},\ldots, x_{i,7})$ containing the value of each of the seven variables in the dataset, as described in Table \ref{tab:case1_vardescription}, and $y_i$ the price of that property. Let the value $x_{i,7}$ be the latitude-longitude coordinates of the location of property $i$. When using a spatial embedding model, we replace the latitude-longitude coordinates of each property with the spatial embedding vector representing those coordinates, constructed by the location encoder $f^\mathrm{LE}$ from a trained spatial embedding model defined in Section \ref{sec:hyperparam}.

\subsubsection{Predictive performance of spatial embeddings}\label{sec:fr_predmodels}

We fit a regression model on the French real estate dataset with the transaction price as the response variable and the variables in Table \ref{tab:case1_vardescription} as input features. We will compare the effect on predictive accuracy when using the latitude and longitude coordinates as input features versus replacing latitude-longitude with its spatial embedding vector. We construct a generalised linear model (GLM), a generalised additive model (GAM), and a gradient boosting machine (GBM). The dataset is split into a randomly selected $80\%$ of the observations as training data and the remaining $20\%$ as an out-of-sample test set.

\added{The embedding models in Table \ref{tab:hyperparam} are general-purpose models that are not fitted to the real estate dataset, and their construction is independent of the response variable. The contrastive pretraining of Section \ref{sec:model_archi} aligns coordinates with spatial views and never observes a transaction price. The pretrained location encoder is a fixed transformation that is applied identically to the training and the test observations, comparable to any deterministic feature engineering step. This set-up is analogous to common practice in machine learning, where convolutional networks pretrained on ImageNet \citep{deng2009imagenet} or word embeddings such as word2vec \citep{mikolov2013efficient} are reused as feature extractors for downstream tasks. The $80/20$ split of the real estate data therefore provides a valid out-of-sample evaluation of the pricing models.}

We fit the GLM \added{on the log-transformed transaction price, using a Gaussian distribution with the identity link function. This specification corresponds to a lognormal regression model for the price itself, in line with the right-skewed price distribution in Figure \ref{fig:case1_responsevar}.} The categorical input features are dummy-encoded for use in the GLM. The levels with the largest amount of exposure are taken as the reference level: $\texttt{type}_{1}$ and $\texttt{land\_use}_{S}$. The location variable is used as two continuous variables, \texttt{latitude} and \texttt{longitude}, and the date variable is used as three continuous variables, \texttt{year}, \texttt{month}, and \texttt{day}. Variable selection is done via a bidirectional stepwise variable selection process; see \citet{Freidmanetal2001}. The full model structure of the GLM using latitude and longitude as input variables is as follows:
\begin{align}
    \added{\mathbb{E}\left(\log\left(\texttt{price}\right)\right)} 
    = &\ \beta_{0} 
    + \beta_{1}\,\texttt{type}_{2}
    + \beta_{2}\,\texttt{type}_{3}
    + \beta_{3}\,\texttt{land\_use}_{\text{AB}}
    + \ldots
    + \beta_{28}\,\texttt{land\_use}_{\text{VI}} \notag \\
    &+ \beta_{29}\,\texttt{rooms}
    + \beta_{30}\,\texttt{building\_area}
    + \beta_{31}\,\texttt{land\_area} \notag \\
    & + \beta_{32}\,\texttt{year}
    + \beta_{33}\,\texttt{month}
    + \beta_{34}\,\texttt{day}
    + \beta_{35}\,\texttt{latitude}
    + \beta_{36}\,\texttt{longitude}.
    \label{eq:glmspecs}
\end{align}

For the GLM using the spatial embedding as input variable, the terms $\beta_{35}\,\texttt{latitude}+\beta_{36}\,\texttt{longitude}$ are replaced with $\beta_{35}\,\texttt{spatial\_embedding}_1+\ldots+\beta_{34+d}\,\texttt{spatial\_embedding}_d$, where $(\texttt{spatial\_embedding}_1,\ldots,\texttt{spatial\_embedding}_d)$ is the spatial embedding vector of dimension $d$. 

We use a Gaussian GAM\added{, fitted on the log-transformed transaction price with the identity link function as well}. The categorical variables are used similarly to those in the GLM, with dummy encoding and the same reference levels. For the continuous variables $\texttt{rooms}$, $\texttt{building\_area}$, and $\texttt{land\_area}$, univariate smooth functions $f_1$, $f_2$, $f_3$ are used. The variable $\texttt{date}$ is treated as three continuous variables, \texttt{year}, \texttt{month}, and \texttt{day}, each with a univariate smooth function $f_4$, $f_5$, $f_6$. The variable $\texttt{coordinates}$ is used as two continuous variables, \texttt{latitude} and \texttt{longitude}, with a bivariate smooth function $f_7$. \added{All smooth functions are fitted with penalised cubic B-splines \citep{eilers1996flexible}, fitted with the pyGAM library. The univariate smooth functions use $20$ basis functions, and the bivariate smooth function on the coordinates is a tensor product of two marginal cubic B-spline bases with $10$ basis functions each. The smoothing parameters are selected through a grid search minimising the generalised cross-validation criterion. The GAM with this bivariate smooth effect on the coordinates serves as a classical spatial smoothing benchmark against which the embedding based models are compared.} The full model structure of the GAM is as follows:
\begin{align}
    \added{\mathbb{E}\left(\log\left(\texttt{price}\right)\right)} 
    = &\ \beta_{0} 
    + \beta_{1}\,\texttt{type}_{2}
    + \beta_{2}\,\texttt{type}_{3}
    + \beta_{3}\,\texttt{land\_use}_{\text{AB}}
    + \ldots
    + \beta_{28}\,\texttt{land\_use}_{\text{VI}} \notag \\
    &\quad + f_{1}(\texttt{rooms})
    + f_{2}(\texttt{building\_area})
    + f_{3}(\texttt{land\_area}) \notag \\
    &\quad + f_{4}(\texttt{year})
    + f_{5}(\texttt{month})
    + f_{6}(\texttt{day})
    + f_{7}(\texttt{latitude},\,\texttt{longitude}).
    \label{eq:gamspecs}
\end{align}

For the GAM using the spatial embedding as input variable, the term $f_{7}(\texttt{latitude},\,\texttt{longitude})$ is replaced by $f_{7}(\texttt{spatial\_embedding}_1) + \ldots+f_{6+d}(\texttt{spatial\_embedding}_d)$.

For the GBM, we use the XGBoost implementation \citep{Chen2016-qm}. The categorical input features are one-hot encoded for use in the GBM. \added{The GBM is fitted on the log-transformed transaction price with a squared error loss, in line with the assumptions made for the GLM and the GAM.} The optimal tuning parameter values are determined via five-fold cross-validation. We tune the number of trees from the values $\{100,500,1000,5000\}$, the depth of each tree from the values $\{3,5,7\}$, and the learning rate from the values $\{0.01,0.001\}$. We assess model performance for the GLM, GAM, and GBM using the mean squared error (MSE) of log-transformed house prices on the test set. The same test set is used for all three models to ensure a fair comparison.

Table \ref{tab:case1_mse} shows the out-of-sample losses for the GLM, GAM, and GBM. The first column specifies how the location of the property was used in the model: either using latitude-longitude as input features (no spatial embedding), or a spatial embedding vector as an input feature, in which case the name of the embedding model follows the acronyms given in Table \ref{tab:hyperparam}. We indicate the lowest out-of-sample loss for each column in bold. We observe that for the GLM, GAM, and GBM, using spatial embeddings instead of latitude-longitude coordinates as input features results in improved out-of-sample loss. The larger the dimension of the embeddings, the lower the out-of-sample loss, which is to be expected as larger embedding dimensions allow for more spatial information to be stored in the embeddings. 

\added{The embeddings improve all three models, but the improvement is more pronounced for the GLM and the GAM. Replacing the coordinates with the \texttt{EU64\_GS64} embedding reduces the out-of-sample loss by approximately $21\%$ for the GLM and $25\%$ for the GAM, and by approximately $7\%$ for the GBM. These percentages are not directly comparable. The GBM with coordinate inputs already attains a lower out-of-sample loss than the GLM and the GAM achieve with spatial embeddings. It constructs a flexible spatial effect by repeatedly splitting on latitude and longitude, and extracts interactions between location and the other covariates from the large training set. Any further reduction is therefore harder to obtain. The embeddings still improve the out-of-sample loss of the GBM, resulting in the lowest value in Table \ref{tab:case1_mse}.}

\added{The row OSM categories in Table \ref{tab:case1_mse} adds the counts of the six OpenStreetMap categories in the central H3 hexagon of each property as direct predictors, next to the latitude and longitude coordinates. This baseline isolates the effect of giving the models direct access to the raw contextual spatial variables. The improvement over the coordinate only models is marginal, and every embedding model outperforms this baseline. The embeddings thus capture spatial structure beyond the local category counts by synthesising information from the surrounding hexagons and from the satellite imagery. The embedding workflow is also simpler in practice, since it only requires coordinates as input, without choosing an H3 resolution and extracting tag counts per observation.}

\added{The row AlphaEarth in Table \ref{tab:case1_mse} uses the pretrained $ 64$-dimensional embeddings of the AlphaEarth foundation model \citep{brown2025alphaearthfoundationsembeddingfield}, extracted from the annual satellite embedding collection in Google Earth Engine and matched to the transaction year of each property. They improve over the raw coordinates for the GLM and the GAM, but not for the GBM. Our \texttt{EU64\_GS64} model has the same dimension and has a lower out-of-sample deviance for all three model classes. AlphaEarth encodes the earth surface for general purpose use at global scale, whereas our framework encodes spatial views chosen for the application at hand, focussed on the European continent.}

\begin{table}[ht!]
    \centering
    \begin{NiceTabular}{lccc}[
    code-before = \rowcolor[HTML]{FFFFFF}{1,3,5,7}
                  \rowcolor[HTML]{FFFFFF}{2,4,6}
    ]
    \toprule
    Spatial treatment & GLM & GAM & GBM \\
    \noalign{\hrule height 0.7pt}
    No spatial embedding          &  $0.3122$ & $0.2362$ & $0.1415$ \\
    \added{OSM categories}        & \added{$0.3099$} & \added{$0.2330$} & \added{$0.1382$} \\
    \added{AlphaEarth}        & \added{$0.2576$} & \added{$0.2306$} & \added{$0.1754$} \\
    \noalign{\hrule height 0.7pt}
    \texttt{EU8\_GS32\_OSM32}     &  $0.3070$ & $0.2147$ & $0.1344$ \\
    \texttt{EU16\_GS32\_OSM16}    &  $0.2861$ & $0.1978$ & $0.1332$ \\
    \texttt{EU16\_OSM16}          &  $0.2775$ & $0.1929$ & $0.1328$ \\
    \texttt{EU32\_GS96\_OSM32}    &  $0.2646$ & $0.1829$ & $0.1325$ \\
    \texttt{EU64\_GS64}           &  $\mathbf{0.2473}$ & $\mathbf{0.1762}$ & $\mathbf{0.1310}$ \\
    \bottomrule
    \end{NiceTabular}
    \caption{Comparison between models using latitude–longitude and spatial embedding as input features. Results show out-of-sample losses, measured in mean-squared error on the log transform of the transaction price. For spatial treatment, the model acronyms are used as defined in Table~\ref{tab:hyperparam}. The lowest out-of-sample loss per column is indicated in bold.}
    \label{tab:case1_mse}
\end{table}

\subsubsection{Model interpretation of spatial effect}\label{sec:fr_interpret}

\begin{figure}[ht!]
    \centering
    \includegraphics[width=1\linewidth]{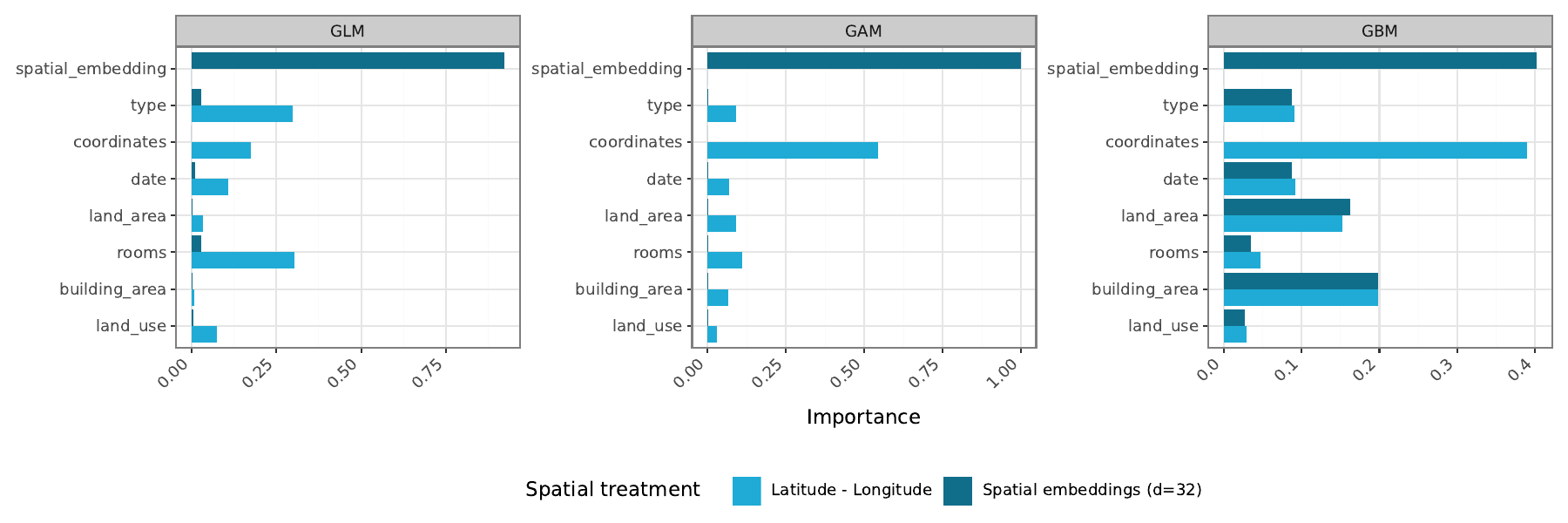}
    \caption{Variable importance as measured in the GLM (left side), the GAM (middle), and the GBM (right side). In dark blue, we show the importance of each variable as measured in the model with spatial embeddings. In light blue, we show the importance measured in the model using latitude-longitude coordinates.}
    \label{fig:case1_vip}
\end{figure}

\paragraph{Variable importance} We use the permutational variable importance tool from \citet{Olden2004} to compare the importance of each variable in our dataset on the prediction of real estate price. The variable importance is measured as the average change in prediction when randomly permuting the values of an input feature. For a prediction model $f_\mathrm{pred}(\cdot)$, and an input feature $x_j$ in the dataset $\mathcal{D}^{\mathrm{FR}}$, the importance of that feature is defined as 
\begin{equation}\label{eq:vip_latlong}
    \mathrm{VIP}_{x_j}= \frac{1}{|\mathcal{D}^{\mathrm{FR}}|}\sum_{\boldsymbol{x}_i\in\mathcal{D}^{\mathrm{FR}}} \mathrm{abs}\big(f_\mathrm{pred}\left(x_{i,1},\ldots,x_{i,j}, \ldots x_{i,7}\right) - f_\mathrm{pred}\left(x_{i,1},\ldots,\tilde{x}_{i,j}, \ldots x_{i,7}\right)\big),
\end{equation}

where $\tilde{x}_{i,j}$ is a random permutation of all observed values of that input variable. When calculating the variable importance of the property’s location, the latitude–longitude pairs are permuted together rather than the latitude and longitude values separately. Similarly, when using the spatial embedding to represent location, the entire embedding vectors are permuted, not the individual elements within each vector. We calculate the variable importance for each of the input variables in Table \ref{tab:case1_vardescription}. Figure \ref{fig:case1_vip} shows the variable importance for the GLM on the left, the GAM in the middle, and the GBM on the right. Dark blue shows the effect of the model with spatial embeddings, whereas light blue shows the effect of the model using latitude-longitude coordinates. For the spatial embedding, we utilised the location encoder from the trained \texttt{EU32\_GS96\_OSM32} model, as described in Section \ref{sec:hyperparam}. We observe in the GLM that the coordinates of the property are the third most important variable. This is because the GLM can only fit two coefficients, one for latitude and one for longitude, which is not sufficient to capture any nuanced spatial effects. Using spatial embeddings, however, the embedding vector is seen as the most important variable in the model. For both the GAM and the GBM, location is the most important variable, whether used via latitude and longitude coordinates or spatial embeddings. For the GLM and the GAM, the effect of permuting the spatial embeddings is so significant that the relative importance of other variables is no longer visible in Figure \ref{fig:case1_vip}. This is not the case for the GBM, where the importance of non-location variables stays relatively equal between the model using latitude and longitude versus the model using spatial embeddings. 

\paragraph{Partial dependence} We use partial dependence plots \citep{Freidmanetal2001} to examine the relationship between the predicted real estate price and the property's location in the pricing model. For a prediction model $f_\mathrm{pred}(\cdot)$, where input feature $x_{i,7}$ represents the latitude-longitude coordinates, we calculate the partial dependence effect of a location $(\lambda,\phi)$ as

\begin{equation}\label{eq:pdp_pc}
    \mathrm{PD}(\lambda,\phi) = \frac{1}{|\mathcal{D}^{\mathrm{FR}}|} \sum_{\boldsymbol{x}_i\in\mathcal{D}^{\mathrm{FR}}} f_\mathrm{pred}(x_{i,1},\ldots,x_{i,6},(\lambda,\phi)).
\end{equation}

When using a prediction model with the spatial embedding of latitude-longitude coordinates as input feature, as calculated via the encoder $f^\mathrm{LE}$, the partial dependence effect of the location  $(\lambda,\phi)$ is calculated as

\begin{equation}\label{eq:pdp_pc_embed}
    \mathrm{PD}(\lambda,\phi) = \frac{1}{|\mathcal{D}^{\mathrm{FR}}|} \sum_{\boldsymbol{x}_i\in\mathcal{D}^{\mathrm{FR}}} f_\mathrm{pred}(x_{i,1},\ldots,x_{i,6},f^\mathrm{LE}(\lambda,\phi)).
\end{equation}

Let $\{(\lambda_c,\phi_c),\, c=1,\ldots,6\,048\}$ denote the latitude–longitude coordinates of the centroids of the $6\,048$ postal codes in France. For each coordinate pair, we compute the partial dependence effect, which results in a spatial map showing how location influences the predicted real estate price.

\begin{figure}[ht]
    \centering
    \includegraphics[width=1\linewidth]{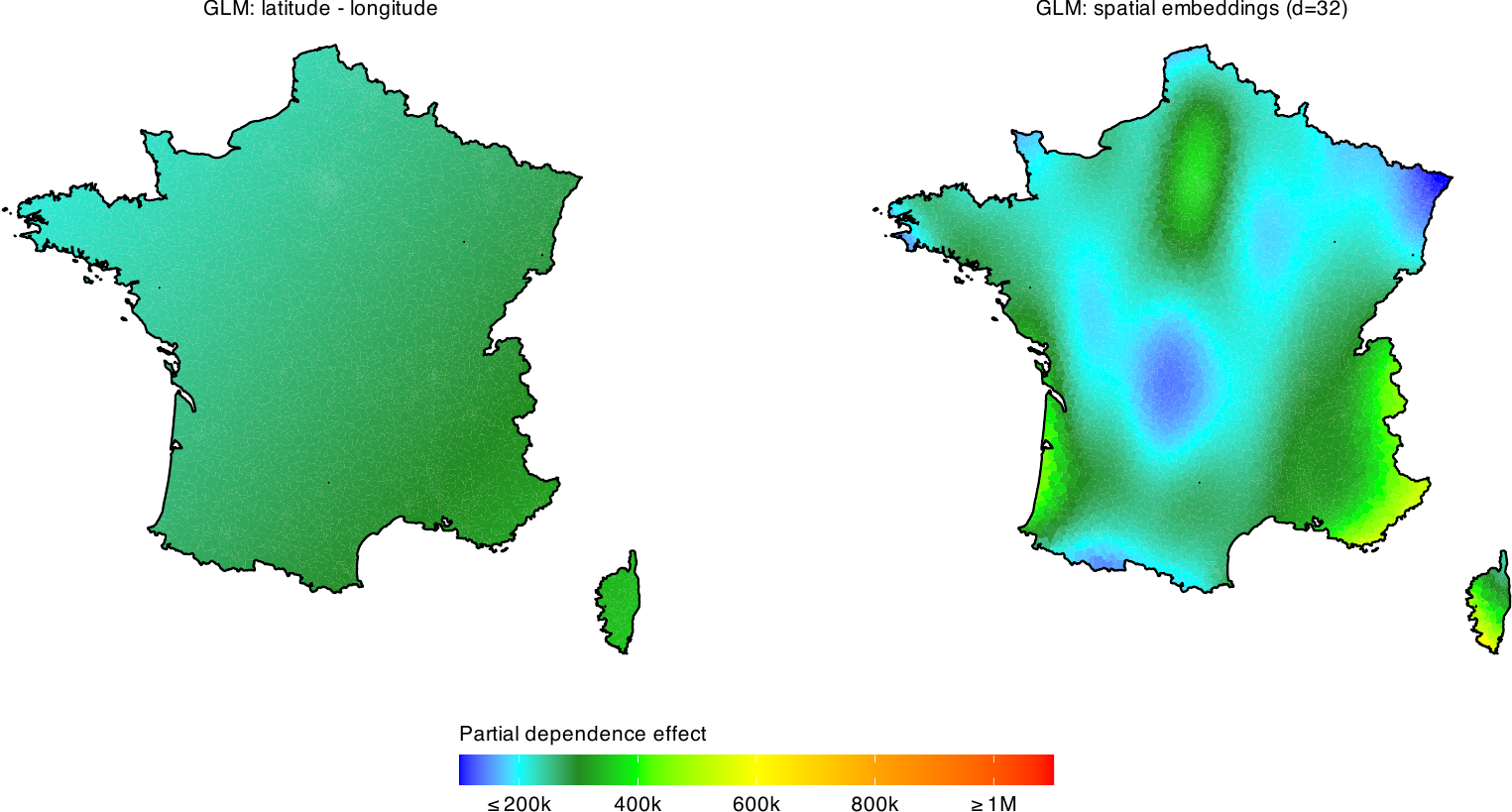}
    \caption{Partial dependence effect of location on real estate price, calculated per postal code for predictions from the GLMs. The left-hand side is the model using latitude-longitude coordinates as input features, and the right-hand side is the model using spatial embeddings instead of latitude-longitude. We use the \texttt{EU32\_GS96\_OSM32} model to extract spatial embeddings.}
    \label{fig:FRhouse_PDP_GLM_comp}
\end{figure}

Figure \ref{fig:FRhouse_PDP_GLM_comp} shows the partial dependence effect as calculated with the GLM. The left-hand side shows the effect of the GLM using latitude and longitude as input features as defined in Equation \eqref{eq:glmspecs}; the right-hand side shows the GLM using the spatial embedding vector as input features. The GLM using latitude and longitude shows a negligible partial dependence effect, with each postal code having a very similar average predicted price. This is in line with the variable importance plots in Figure \ref{fig:case1_vip}, which indicate that the variables latitude and longitude are not important in this GLM. By contrast, the partial dependence effect for the GLM using spatial embeddings shows higher predicted prices around Paris and in the south-east and south-west regions of France, with a further effect visible on the island of Corsica. This demonstrates that the GLM can capture more nuanced patterns through spatial embeddings than it can solely based on latitude and longitude coordinates. 

\begin{figure}[ht]
    \centering
    \includegraphics[width=1\linewidth]{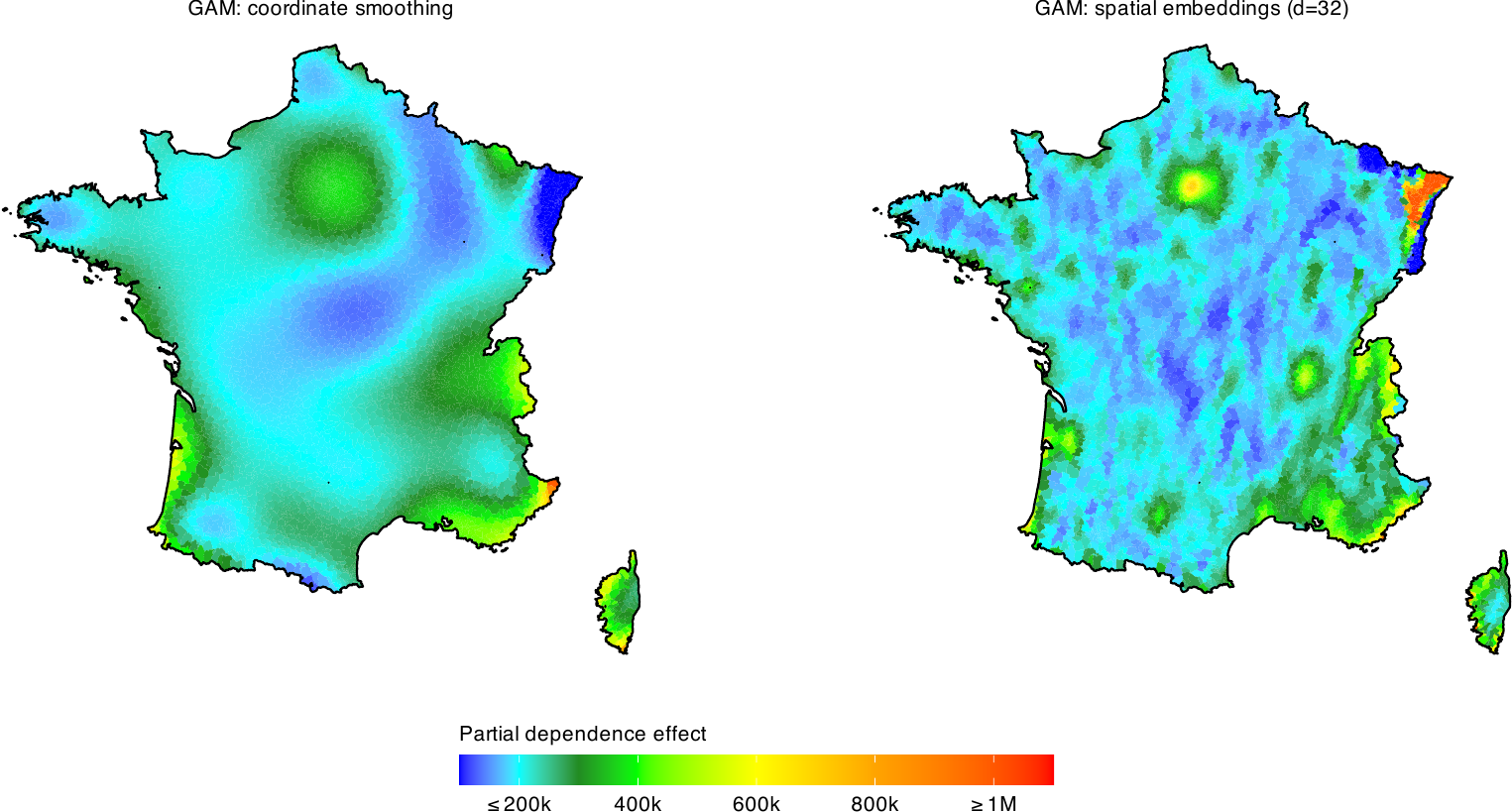}
    \caption{Partial dependence effect of location on real estate price, calculated per postal code for predictions from the GAM. The left-hand side is the model using latitude-longitude coordinates as input features, and the right-hand side is the model using spatial embeddings instead of latitude-longitude. We use the \texttt{EU32\_GS96\_OSM32} model to extract spatial embeddings.}
    \label{fig:FRhouse_PDP_GAM_comp}
\end{figure}

Figure \ref{fig:FRhouse_PDP_GAM_comp} shows the partial dependence effect per postal code as calculated with the GAM. The left-hand side displays the GAM using latitude and longitude as input, as described in Equation \eqref{eq:gamspecs}. Compared to the GLM, the bivariate smoothing on the coordinates produces a more nuanced spatial pattern, with higher predicted prices around Paris and in the south-east and south-west regions. The right-hand side shows the GAM using spatial embeddings as input. In this case, the model captures more complex spatial variation, distinguishing between urban centres and rural areas.

\begin{figure}[ht]
    \centering
    \includegraphics[width=1\linewidth]{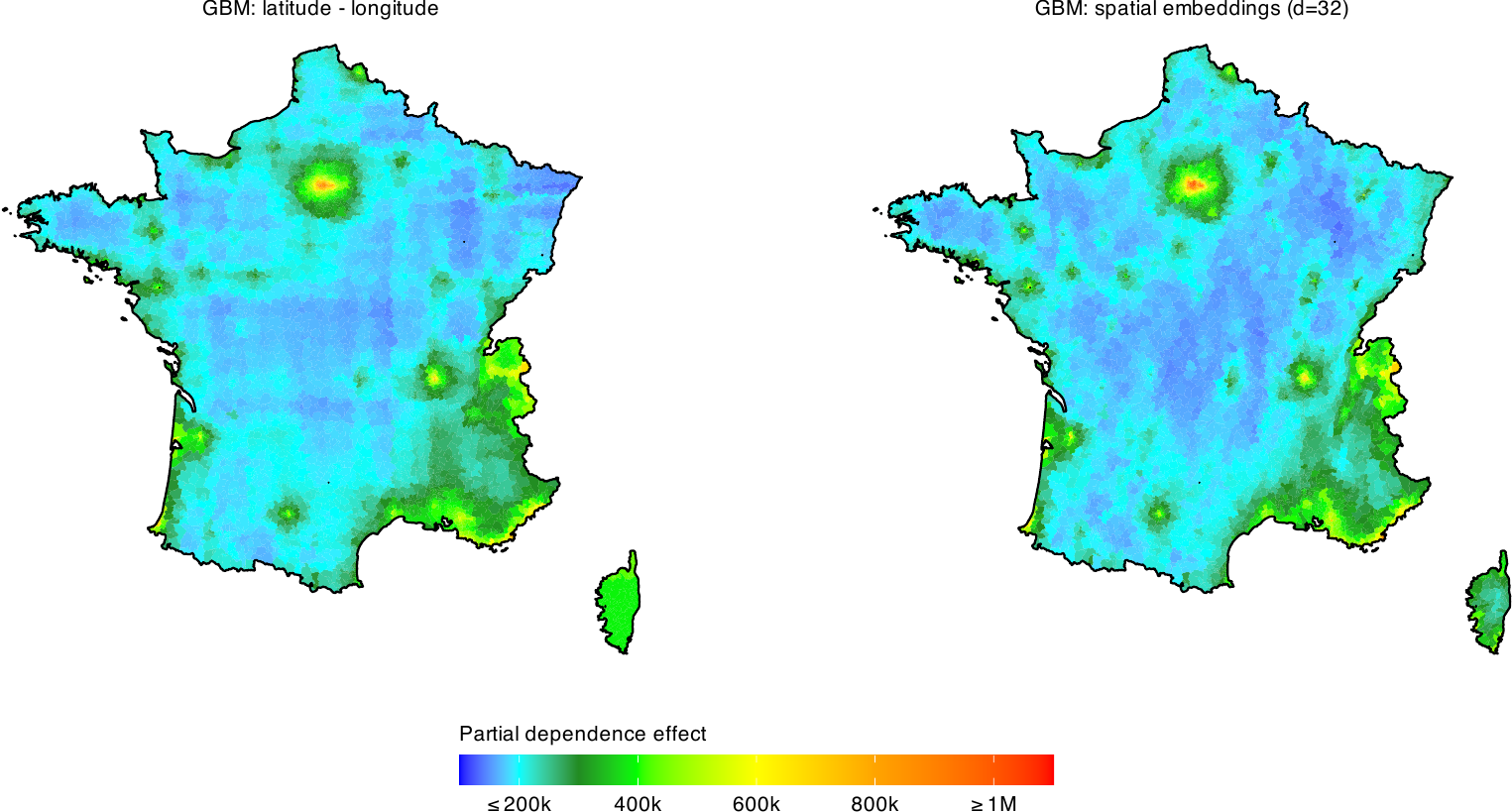}
    \caption{Effect of location on average predicted real estate price, calculated per postal code for predictions from the GBMs. The left-hand side is the model using latitude-longitude coordinates as input features, and the right-hand side is the model using spatial embeddings instead of latitude-longitude. We use the EU32\_GS96\_OSM32 model to extract spatial embeddings.}
    \label{fig:FRhouse_PDP_GBM_comp}
\end{figure}

Figure \ref{fig:FRhouse_PDP_GBM_comp} shows the partial dependence effect of the GBM using latitude-longitude as input features on the left and using the spatial embeddings on the right-hand side. For both the GBM using latitude-longitude and the GBM using spatial embeddings, we see a difference between cities in France and rural areas. The GBM using latitude-longitude, however, shows a square pattern caused by the linear splits made by the regression trees in the GBM \citep{henckaerts2020boosting, Geerts2024-dk}. The partial dependence effect from the GBM using spatial embeddings does not show these square artifacts and shows the shapes of cities more clearly. \added{At the scale of Figure \ref{fig:FRhouse_PDP_GBM_comp}, both GBMs thus recover the dominant spatial pattern, with higher predicted prices in and around the cities and lower predicted prices in the rural centre. This similarity breaks down at finer spatial scales. The splits on latitude and longitude divide the map into rectangular regions with a constant predicted effect, so the coordinate-based GBM cannot represent spatial variation within such a region.}

\paragraph{Fine-grained partial dependence effects}

Instead of one coordinate pair per postal code, we now define a fine-grained set of locations in the area between Lyon and Switzerland. We randomly sample a set of $10\,000$ locations in this area. By calculating the partial dependence effect for this set of locations, we identify fine-grained effects in the relationship between location and predicted prices. We calculate the partial dependence effect for each location following Equation \eqref{eq:pdp_pc_embed} using the GBM. 

\begin{figure}[ht]
    \centering
    \includegraphics[width=\linewidth]{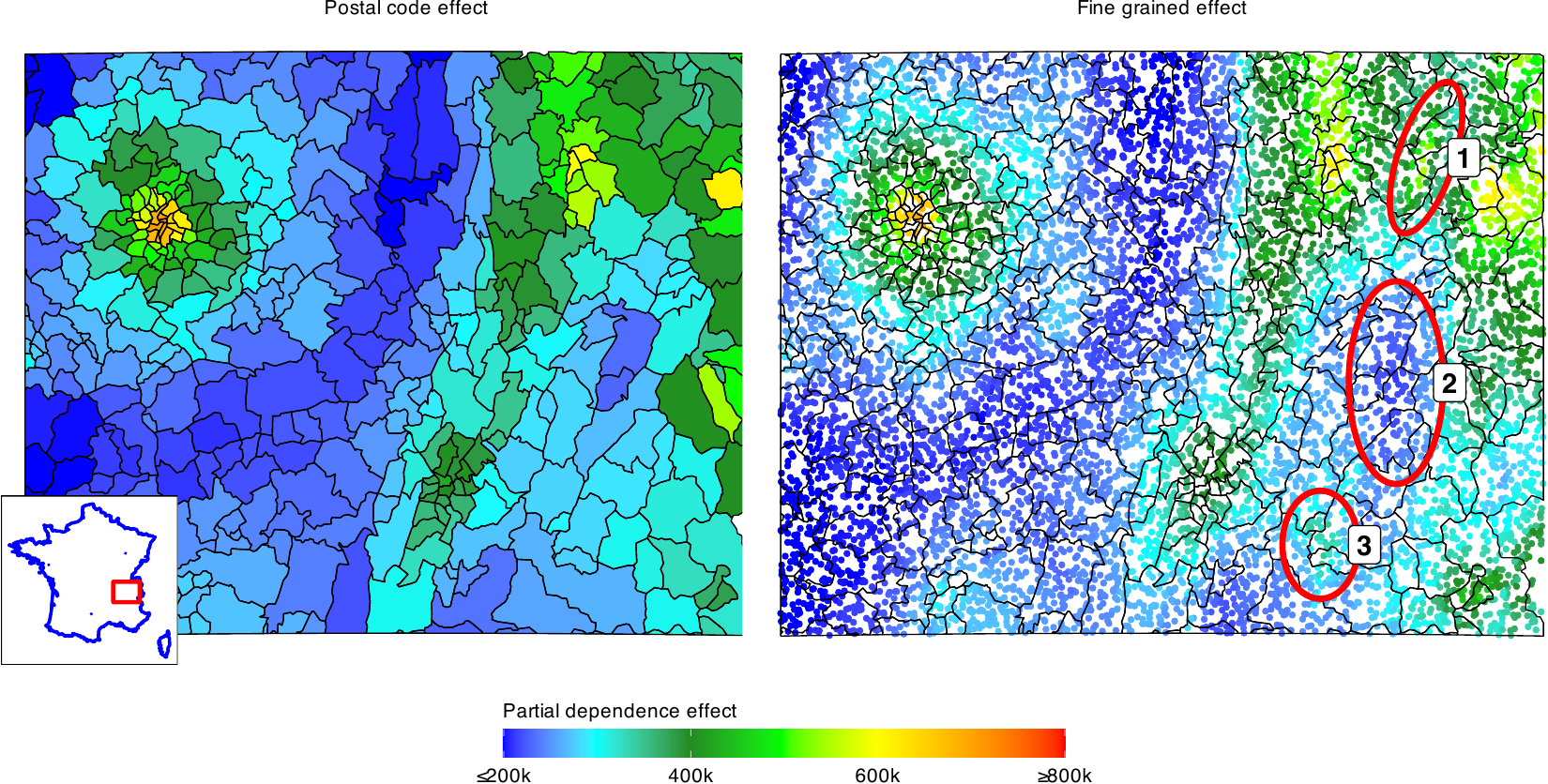}
    \caption{Partial dependence effect from the GBM using spatial embeddings calculated in a selected area of France between Lyon and Switzerland. The used spatial embedding model is the \texttt{EU32\_GS96\_OSM32} model. The selected area is indicated on the small map of France on the bottom left. The left-hand side shows the partial dependence effect calculated on the level of postal codes, and the right-hand side shows the partial dependence effect calculated on $10\ 000$ randomly sampled points in the selected area.}
    \label{fig:FRhouse_finegrained_pdp}
\end{figure}

Figure \ref{fig:FRhouse_finegrained_pdp} shows the partial dependence effects. The left-hand side displays results at the postal code level, while the right-hand side shows the fine-grained set of locations. Three localised patterns appear in the fine-grained map that are not visible at the postal code level, highlighted by red ovals. The first is a small, oblong area with higher predicted prices, corresponding to the touristically popular northwestern slopes of the Aravis range. The second is a wide oblong area with lower predicted prices, coinciding with the La Lauzière range. The La Lauzière mountain range is less touristic and houses only a few resorts or ski areas. The third is an angular pattern of higher predicted prices matching the Les Trois Vallées area. This is the largest connected ski area in the world and is highly touristic. These examples demonstrate that the GBM has fitted a relationship between real estate prices and the spatial information captured in the spatial embeddings.

\paragraph{Generalisation of spatial effects to regions without training observations}

\added{We examine how the spatial effect fitted by the GBM generalises to regions without price observations in the training data. The departments Bas-Rhin, Haut-Rhin, and Moselle are absent from the French real estate dataset, since transactions in these departments are not reported to the DVF database. We compare the partial dependence of the coordinate-based and the embedding-based GBM for postal codes located in these departments. The embedding model has observed the region during its pretraining, since the spatial views are available there, but the pricing model has never observed any property there.}

\begin{figure}[htb]
    \centering
    \includegraphics[width=1\linewidth]{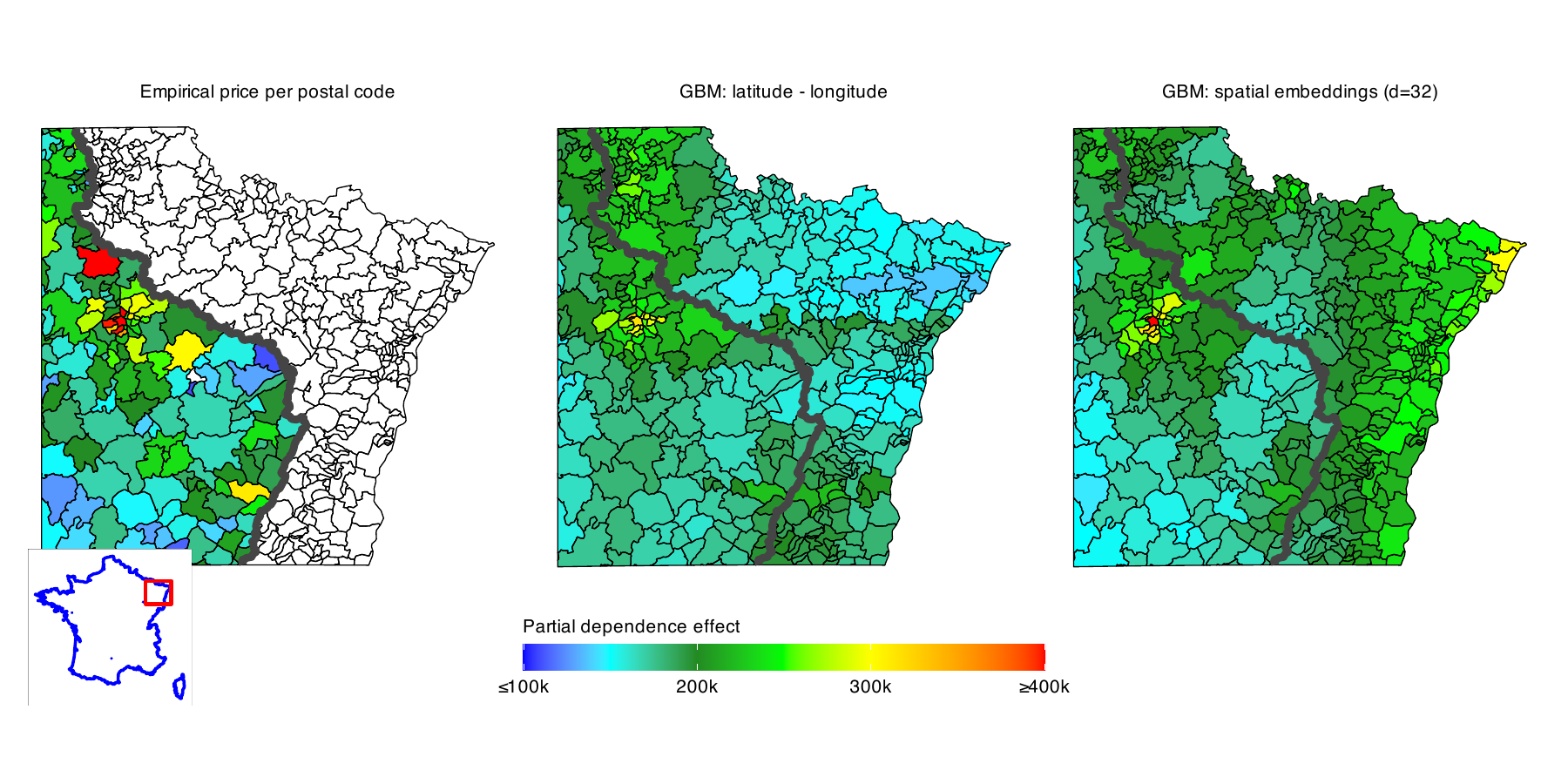}
    \caption{Selected area in France between the city of Nancy and the German border, as indicated on the inset map (bottom left). The left panel shows the empirically observed average transaction price per postal code, where white corresponds to postal codes without observations in the dataset, namely in the departments Bas-Rhin, Haut-Rhin, and Moselle. The middle panel shows the partial dependence effect calculated with the GBM using latitude and longitude as input. The right panel shows the partial dependence effect from the GBM using spatial embeddings. The grey line marks the boundary between regions included in the training data and regions without observations.}
    \label{fig:FRhouse_transfer_pdp}
\end{figure}

The first panel of Figure \ref{fig:FRhouse_transfer_pdp} shows the empirical mean transaction price per postal code in the region between Nancy and the German/Swiss border; a map of France in the lower left highlights the selected area. A thick grey line marks the spatial extent of the training data, and postal codes in white indicate areas absent from the dataset. The second panel shows the partial dependence (PD) for this region obtained from the GBM using latitude and longitude as inputs. Because the excluded departments lie beyond the range of latitude and longitude values present in the training data, the GBM cannot form new splits in these areas. Instead, the model extrapolates by assigning these unseen locations to the nearest regions learnt during training, resulting in an unrealistic, piecewise-constant partial dependence effect. The third panel shows the partial dependence effect obtained from the GBM using spatial embeddings as input features. Here, the model captures a lower partial dependence effect in the central part of the region and a higher effect along the German border. Both patterns are intuitive: the central area is predominantly rural with lower population density, whereas larger cities, such as Strasbourg, are located along the border. By using spatial embeddings, the model associates predictions with spatial characteristics rather than with absolute coordinates. \added{This representation allows the fitted spatial effect to generalise to regions without training observations by exploiting similarities in spatial structure. The partial dependence pattern, rather than the absolute price level, is validated here, since no transaction prices are available in the excluded departments.}

\subsection{Insurance case study: flood claim counts in Belgium}\label{sec:flood_casestudy}

\leavevmode\added{We perform a second case study on insurance claim counts, using the flood claims recorded in the Klimaatschademonitor \citep{assuralia2025klimaatschademonitor} of Assuralia, the federation of Belgian insurance companies. The dataset contains the number of claims per Belgian postal code and per calendar year, for the period $2015$ to $2025$ and separately for the perils vehicle damage, flood, and storm. We focus on the flood peril. Insurance datasets with claims linked to policies at the coordinate or address level are proprietary, but this open dataset allows us to construct a reproducible insurance case study. The dataset contains $1\,146$ postal codes and $12\,606$ postal code and calendar year combinations. The members of Assuralia represent nearly the entire Belgian insurance market, and flood damage coverage is a mandatory part of Belgian fire insurance policies. Hence, the recorded claims effectively span the insured building stock of the country. There is no exposure measure in the dataset, such as the number of policies or insured buildings per postal code. The models therefore estimate the expected number of flood claims per postal code and calendar year, rather than a claim frequency per unit of exposure, and the spatial variation they capture reflects both the concentration of insured buildings and the underlying flood risk.}

\begin{figure}[ht!]
    \centering
    \begin{subfigure}[b]{0.32\linewidth}
        \includegraphics[width=\linewidth,height=1.465\linewidth,keepaspectratio]{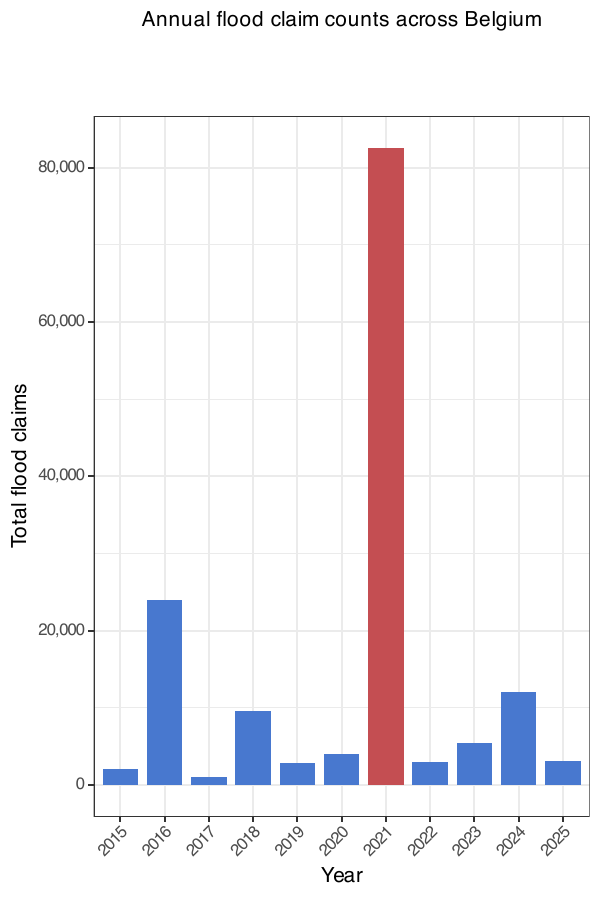}
        \caption{\added{Total number of flood claims per calendar year in Belgium.}}
        \label{fig:BE_flood_bar}
    \end{subfigure}
    \hfill
    \begin{subfigure}[b]{0.64\linewidth}
        \includegraphics[width=\linewidth]{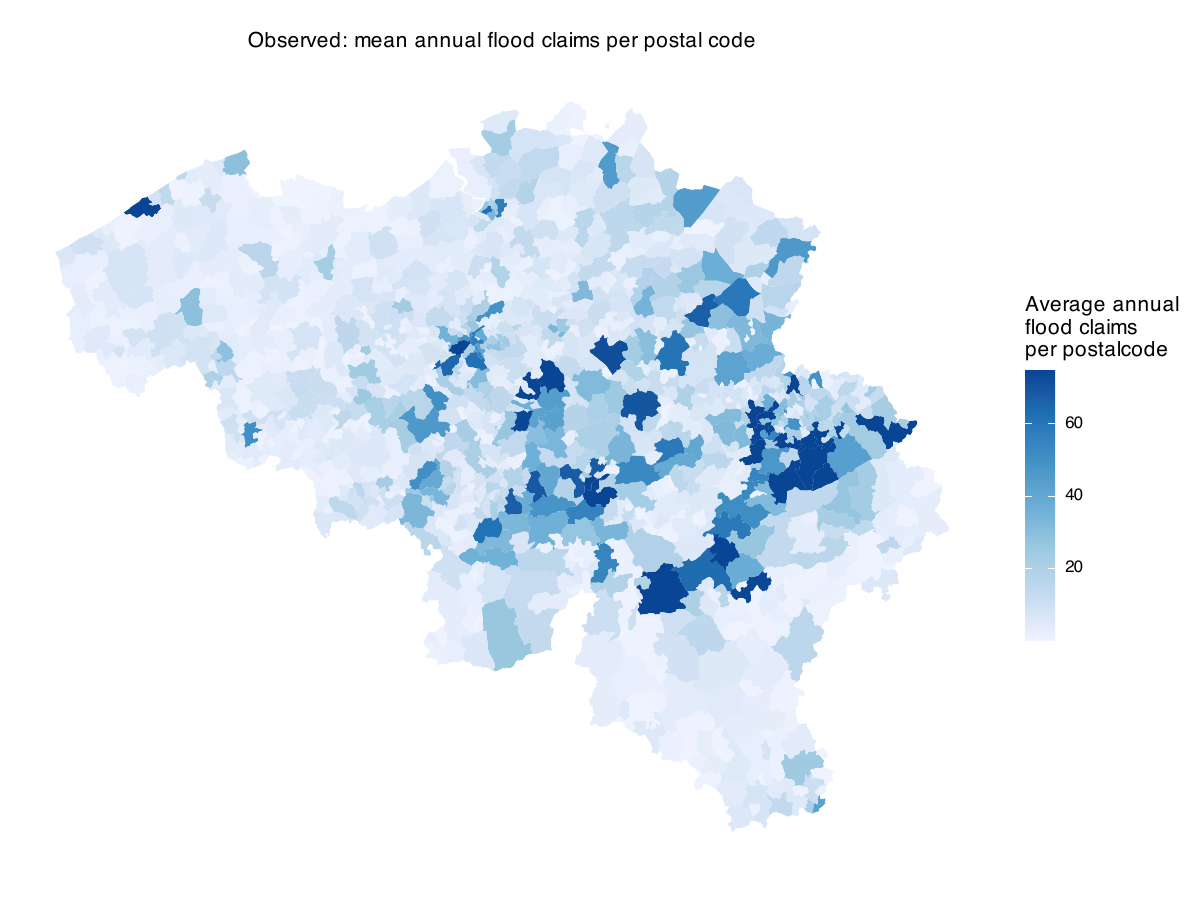}
        \caption{\added{Observed average annual number of flood claims per Belgian postal code.}}
        \label{fig:BE_flood_map}
    \end{subfigure}
    \caption{\added{Flood claims recorded in the Klimaatschademonitor of Assuralia \citep{assuralia2025klimaatschademonitor}, for the period 2015 to 2025. Panel (a) shows the annual totals across Belgium. Panel (b) shows the average annual number of claims per postal code.}}
    \label{fig:BE_flood_data}
\end{figure}

\added{Figure \ref{fig:BE_flood_bar} shows the total number of flood claims per calendar year and Figure \ref{fig:BE_flood_map} the observed average annual number of claims per postal code. The annual claim counts per postal code are heavily right-skewed. About $35\%$ of the observations record no flood claims, the average count per year and per postal code is $11.9$ claims. The maximum observed number is $3\,401$ claims, recorded in postal code 4800 (Verviers) during the July $2021$ floods in the east of Belgium. The year $2021$ alone accounts for $55\%$ of all flood claims in the dataset, more than three times the total of any other year, as seen in Figure \ref{fig:BE_flood_bar}. The spatial pattern in Figure \ref{fig:BE_flood_map} concentrates along the Meuse and Vesdre valleys in the east of the country.}

\added{We denote the Belgian flood dataset as $\mathcal{D}^{\mathrm{BE}} = \left(\boldsymbol{x}_i, y_i\right)_{i=1}^{12\,606}$, where each index $i$ represents one postal code and calendar year combination, $y_i$ the observed number of flood claims, and $\boldsymbol{x}_i$ the location of the postal code centroid. The location is used either as the latitude and longitude coordinates or as the $32$-dimensional spatial embedding obtained by evaluating the location encoder of the trained \texttt{EU32\_GS96\_OSM32} model at the centroid. The postal code itself is not used as a categorical variable. We compare a Poisson GLM, a Poisson GAM, and a Poisson GBM, all with a logarithmic link function. The GLM uses the latitude and longitude coordinates, or the $32$ embedding components, as linear terms. The GAM uses a bivariate smooth function of latitude and longitude, or a univariate smooth function per embedding component, fitted with penalised cubic B-splines. The GBM uses the XGBoost implementation \citep{Chen2016-qm} with the Poisson deviance as loss function, $500$ trees of depth $4$, a learning rate of $0.05$, and a subsample fraction of $0.8$. The calendar year is not used as a predictor, so each model estimates an average annual number of flood claims per postal code.}

\added{Annual flood claim counts vary strongly between calendar years, driven by individual events such as the July $2021$ floods in Belgium. A temporal train and test split would therefore mainly measure calendar year volatility rather than the quality of the territorial risk classification. We instead use five-fold cross validation grouped by postal code. We assign the $1\,146$ postal codes at random to five sets of approximately equal size. Each set serves once as the hold-out set, while the models are fitted on the postal codes in the other four set. Grouping by postal code means that all calendar year observations of a postal code belong to the same set. Each postal code then gets a prediction by a model that was not fitted on any of its observations, which directly tests the assignment of risk levels to new territories.} \added{We calculate the out-of-sample Poisson deviance, averaged over the five sets, for each model. For a model $f$ and a set of $n$ hold-out observations, the Poisson deviance is calculated as}

\begin{equation}
    \added{D_{\mathrm{Poisson}}\left(f(\boldsymbol{x}), y\right) = \frac{2}{n} \sum_{i=1}^{n} \left( y_i \ln\frac{y_i}{f(\boldsymbol{x}_i)} - \left(y_i - f(\boldsymbol{x}_i)\right) \right),}
    \label{eq:flood_deviance}
\end{equation}

\added{with the convention $0\ln 0$ is put equal to zero for observations without claims.}

\added{Table \ref{tab:flood_deviance} shows the out-of-sample Poisson deviance for the three model classes. The model using spatial embeddings attains a lower deviance than those using coordinates for all three model classes. The GLM with spatial embeddings attains a deviance below that of both the GAM with a bivariate coordinate smoother and the GBM on raw coordinates. The deviance is dominated by a small number of postal codes with extreme claim counts, given the heavy right skew. It measures how close the predicted claim counts are to the observed counts, rather than how well a model separates the high-risk postal codes from the low-risk ones.}

\begin{table}[ht!]
    \centering
    \begin{NiceTabular}{lccc}
    \toprule
    \added{Feature set} & \added{GLM} & \added{GAM} & \added{GBM} \\
    \noalign{\hrule height 0.3pt}
    \added{Coordinates}       & \added{$52.31$} & \added{$49.60$} & \added{$49.99$} \\
    \added{Spatial embedding} & \added{$49.42$} & \added{$49.35$} & \added{$\mathbf{48.88}$} \\
    \bottomrule
    \end{NiceTabular}
    \caption{\added{Out-of-sample Poisson deviance for the flood claim count models, averaged over the five sets of the cross-validation grouped by postal code. The lowest value is indicated in bold. The spatial embedding is extracted with the \texttt{EU32\_GS96\_OSM32} model.}}
    \label{tab:flood_deviance}
\end{table}

\begin{figure}[ht!]
    \centering
    \includegraphics[width=\linewidth]{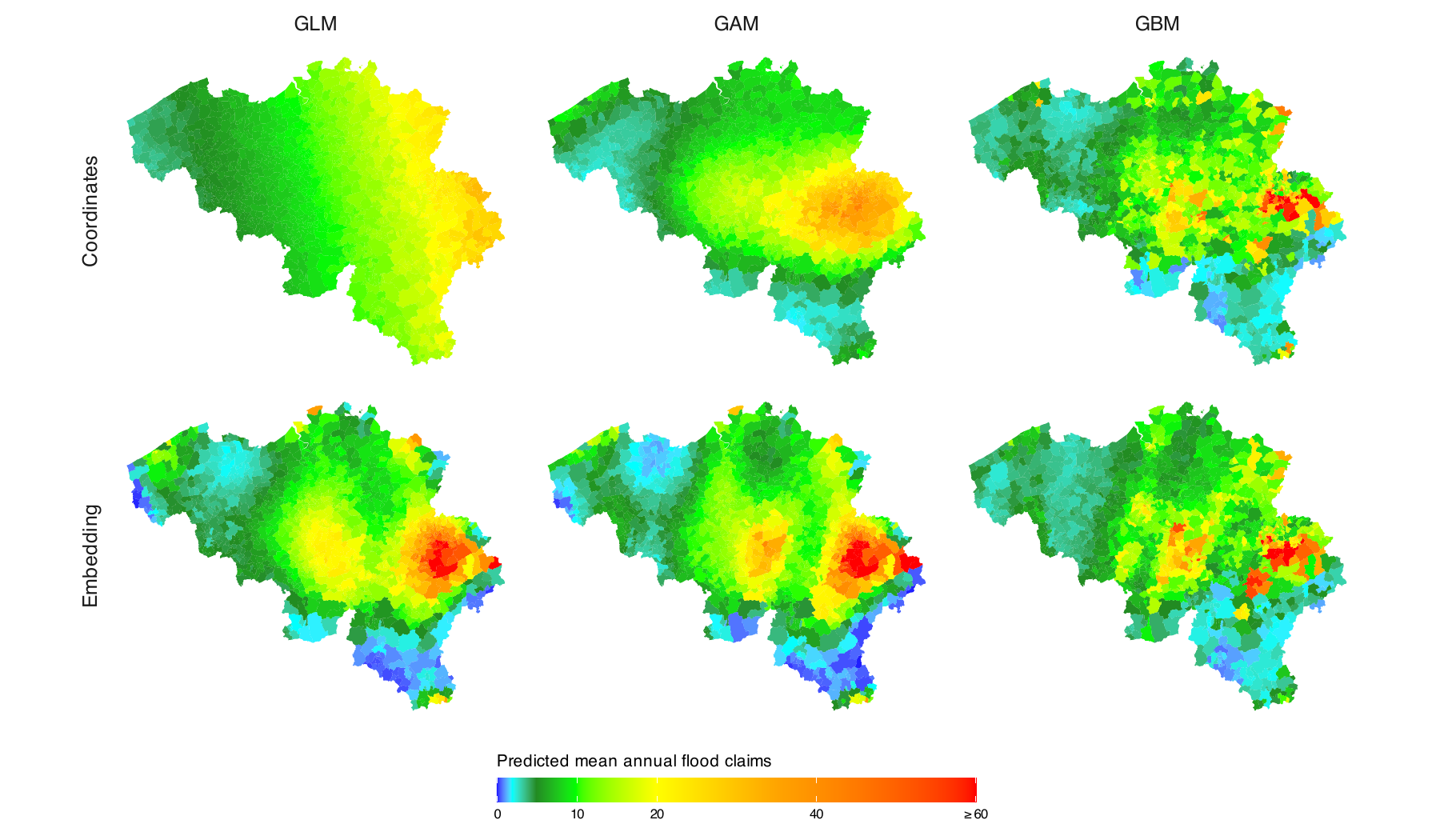}
    \caption{\added{Predicted average annual number of flood claims per Belgian postal code, for the GLM (left), the GAM (middle), and the GBM (right). The top row shows the models using the latitude and longitude coordinates of the postal code centroid, the bottom row the models using the spatial embeddings of that centroid. All predictions are made out of sample. The spatial embeddings are extracted with the \texttt{EU32\_GS96\_OSM32} model.}}
    \label{fig:BE_flood_riskmaps}
\end{figure}

\added{Figure \ref{fig:BE_flood_riskmaps} shows the predicted average annual number of flood claims per postal code. The GLM on the coordinates produces a smooth gradient from west to east, since the two coefficients can only tilt the predicted counts in a single direction. The GAM on the coordinates adds one broad area of higher predicted counts in the east, without separating the postal codes within that area. The GBM on the coordinates distinguishes areas of higher and lower predicted counts, with strong contrasts between neighbouring postal codes. The models using the spatial embeddings separate areas of high and low predicted counts for all three model classes. They locate the highest predicted counts around Verviers and Li\`ege, in line with the observed pattern in Figure \ref{fig:BE_flood_map}. The GLM on the spatial embeddings thus recovers the observed flood pattern more closely than the GAM with a bivariate smoother on the coordinates, which spreads the risk over a single broad region in the east. The GLM and the GAM on the spatial embeddings produce nearly the same map, and both vary more smoothly over neighbouring postal codes than the GBM.}

\added{The model predictions do not depend on the calendar year, so each postal code receives a single predicted annual claim count. We denote by $\hat{y}_p$ the predicted average annual number of flood claims for postal code $p$ under a given model, and by $\bar{y}_p$ the corresponding observed average over the calendar years. Following \citet{holvoet2023neural}, we compare the risk classification achieved by the models using Lorenz curves \citep{lorenz1905methods}. For a model $f$, let $F_n$ be the empirical cumulative distribution function of the predictions $\hat{y}_p$ over the $n = 1\,146$ postal codes, and $r^f_p = F_n(\hat{y}_p)$ the risk score of postal code $p$. The Lorenz curve evaluated in $s \in [0,1]$ is}

\begin{equation}
    \added{\mathrm{LC}^f(s) = \frac{\sum_{p} \bar{y}_p \, \mathbbm{1}\left\{r^f_p \leq s\right\}}{\sum_{p} \bar{y}_p}.}
    \label{eq:flood_lorenz}
\end{equation}

\begin{figure}[ht!]
    \centering
    \includegraphics[width=0.5\linewidth]{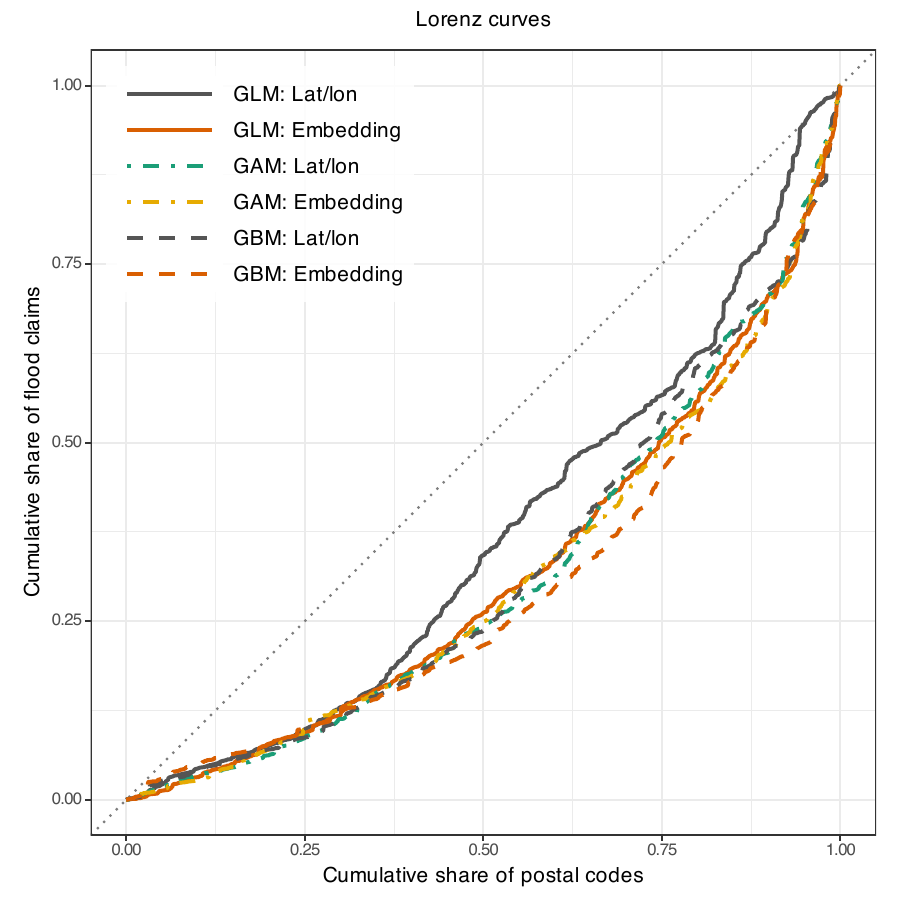}
    \caption{\added{Lorenz curves for the flood claim count models, based on the out-of-fold predictions per postal code. A grey dotted line shows the line of equality.}}
    \label{fig:BE_flood_lorenz}
\end{figure}

\added{The Lorenz curve shows the accumulation of observed claims, ordered by the risk score obtained with model $f$. A curve further away from the line of equality represents a better risk classification. Figure \ref{fig:BE_flood_lorenz} shows the Lorenz curves of the out-of-fold predictions. Within each model class, the curve of the model using the spatial embeddings lies further away from the line of equality than the curve of the model using the coordinates. The GBM using the spatial embeddings lies furthest away, consistent with its lowest out-of-fold deviance in Table \ref{tab:flood_deviance}.}

\added{We use the ordered Lorenz curve, introduced by \citet{frees2011summarizing}, to directly compare the risk classification under one model relative to another. For two models $f_A$ and $f_B$ with predictions $\hat{y}^A_p$ and $\hat{y}^B_p$ for postal code $p$, we define the relativities of model B over model A as}

\begin{equation}
    \added{\mathrm{rel}^{(A,B)}_p = \frac{\hat{y}^B_p}{\hat{y}^A_p}.}
    \label{eq:flood_relativities}
\end{equation}

\added{With $F_n$ the empirical distribution function of the relativities $\mathrm{rel}^{(A,B)}_p$ over all postal codes, the ordered Lorenz curve is defined as}

\begin{equation}
    \added{\left(
    \frac{\sum_{p} \hat{y}^A_p \, \mathbbm{1}\left\{F_n\left(\mathrm{rel}^{(A,B)}_p\right) \leq s\right\}}{\sum_{p} \hat{y}^A_p},
    \frac{\sum_{p} \bar{y}_p \, \mathbbm{1}\left\{F_n\left(\mathrm{rel}^{(A,B)}_p\right) \leq s\right\}}{\sum_{p} \bar{y}_p}
    \right),}
    \label{eq:flood_ord_lorenz}
\end{equation}

\added{over $s \in [0,1]$. The ordered Lorenz curve illustrates the potential improvement in risk classification an insurer can achieve by switching from model $f_A$ to model $f_B$. The Gini index is equal to twice the area between the line of equality and the curve, and a Gini index close to zero indicates that model $f_B$ does not improve on model $f_A$ \citep{frees2011summarizing}. Table \ref{tab:flood_gini} presents the Gini indices for all pairs of models. Following the min-max strategy of \citet{frees2014insurance}, we select the model A with the lowest maximum Gini index, since this model leaves the least potential for improvement in risk classification. This criterion selects the GBM with spatial embeddings, with a maximal Gini index of $15.86$.}

\setlength{\extrarowheight}{3pt}
\begin{table}[ht!]
    \centering
    \small
    \begin{NiceTabular}{cr*{6}{r}}
    \toprule
    & & \Block{1-6}{\footnotesize \added{\textbf{Model B}}} \\[1mm]
    \RowStyle{\rotate}
    & & \added{GLM coordinates} & \added{GLM embedding} & \added{GAM coordinates} & \added{GAM embedding} & \added{GBM coordinates} & \added{GBM embedding} \\
    \noalign{\hrule height 0.3pt}
    \parbox[t]{2mm}{\multirow{6}{*}{\rotatebox[origin=c]{90}{\footnotesize \added{\textbf{Model A}}}}}
    & \added{GLM coordinates} & \phantom{$00.00$} & \added{$29.54$} & \added{$26.84$} & \added{$31.06$} & \added{$29.39$} & \added{$\mathbf{33.38}$} \\
    & \added{GLM embedding}   & \added{$4.61$}  & \phantom{$00.00$} & \added{$4.23$}  & \added{$10.32$} & \added{$15.84$} & \added{$\mathbf{20.47}$} \\
    & \added{GAM coordinates} & \added{$0.09$}  & \added{$14.97$} & \phantom{$00.00$} & \added{$17.54$} & \added{$18.46$} & \added{$\mathbf{22.80}$} \\
    & \added{GAM embedding}   & \added{$5.28$}  & \added{$3.35$}  & \added{$6.55$}  & \phantom{$00.00$} & \added{$13.73$} & \added{$\mathbf{17.29}$} \\
    & \added{GBM coordinates} & \added{$13.18$} & \added{$16.38$} & \added{$16.28$} & \added{$17.25$} & \phantom{$00.00$} & \added{$\mathbf{18.86}$} \\
    & \added{\ulcolor[blue!25]{GBM embedding}}  & \added{$9.80$}  & \added{\cellcolor{blue!25}$\mathbf{15.86}$} & \added{$14.52$} & \added{$14.82$} & \added{$9.96$} & \phantom{$00.00$} \\
    \bottomrule
    \end{NiceTabular}
    \caption{\added{Gini indices, in percent, calculated as twice the area between the ordered Lorenz curve and the diagonal for model B relative to model A, computed on the out-of-fold predictions per postal code. Model A is the benchmark model and model B the competing model, both taken from the six models in Table \ref{tab:flood_deviance}, with the relativities of model B over model A defined in Equation \eqref{eq:flood_relativities}. For each row, the maximum Gini index is given in bold. The min-max strategy selects the model A with the lowest maximum Gini index, shown by the highlighted cell and underlined model.}}
    \label{tab:flood_gini}
\end{table}

\added{The July $2021$ floods dominate the claim totals in Figure \ref{fig:BE_flood_bar} and could in principle drive the rankings on their own. We therefore repeat the full evaluation, including the refitting of all models, on the data without calendar year $2021$, using the same set assignment. Within each model class, the model using the spatial embeddings retains the lower out-of-sample deviance, with $18.68$ against $19.27$ for the GLM, $18.68$ against $18.73$ for the GAM, and $18.34$ against $18.82$ for the GBM. The maximum Gini index per model is lower without this year, which shows that the extreme year carries a substantial part of the spatial signal. The min-max strategy again selects the GBM with spatial embeddings, with a maximal Gini index of $8.77$. The ranking improvements are thus not an artefact of a single extreme event.}

\added{This case study demonstrates the added value of spatial embeddings in an actuarial setting, using open data at the postal code resolution. The embeddings improve the out-of-sample territorial risk classification of the GLM, the GAM, and the GBM on flood claim count modelling. The postal code centroids used in this case study only capture part of the spatial resolution of the embeddings, since a single point represents an entire postal code area. We expect more substantial gains with granular policy-level data within an insurance company, where exact building locations are available and the number of policies per location enters the Poisson models as an exposure offset.}

\section{Conclusion} \label{sec:conclusion}

In this paper, we introduced a framework for constructing spatial embeddings using contrastive learning in a multi-view setting. To demonstrate the multi-view spatial embedding model, we combined satellite imagery and OpenStreetMap tags to learn low-dimensional representations of geographic locations. These embeddings are trained to align with coordinate-based encodings, allowing any dataset with latitude–longitude information to be enriched with spatial context without requiring direct access to the original spatial inputs. The paper is accompanied by a GitHub repository at \url{https://github.com/freekholvoet/MultiviewSpatialEmbeddings}, which provides the code for training the spatial embedding model. The trained models defined in Table \ref{tab:hyperparam} are available in \citet{holvoet_mvmodels_2025}.

We demonstrated the practical value of these embeddings in a real estate pricing application \added{and in an insurance case study on flood claim counts in Belgium}. Replacing raw coordinates with spatial embeddings consistently improved predictive performance across GLMs, GAMs, and GBMs. The embeddings captured spatial structure more effectively than traditional location features and provided \added{post-hoc explainable} spatial effects. Variable importance and partial dependence analyses showed that the embeddings learnt meaningful geographic patterns, including urban–rural differences, regional economic variation, and fine-grained effects related to topography and land use. These effects are difficult to capture using raw coordinates or categorical location indicators such as postal codes. Moreover, the use of spatial embeddings enabled models to generalise \added{their fitted spatial effects} to regions \added{without outcome observations in the training data,} by associating predictions with spatial characteristics rather than coordinates. This property is particularly relevant for insurance applications, where actuaries often need to assess risk in new or underrepresented areas. \added{The real estate case study was chosen deliberately as the primary demonstration. The relationship between location and transaction prices is strong and well understood, which makes the partial dependence effects directly verifiable against known geography, with high prices around Paris and in popular ski areas and lower prices in the rural centre of France. Such verification would be far more difficult on a noisier insurance outcome.} \added{The flood case study in Section \ref{sec:flood_casestudy} complements this verification with direct evidence on insurance data, where the embeddings improve the out-of-sample risk classification of flood claims across Belgian postal codes. Together with existing applications of learnt spatial representations in actuarial science, such as claim frequency modelling for motor and home insurance \citep{blier-wong2022geographic}, these results support the use of spatial embeddings across a broad range of insurance applications. A natural next step is the application of the framework to granular policy level data within an insurance company, where the embeddings can be combined with exposure information and policyholder characteristics.}

From an actuarial modelling perspective, spatial embeddings offer a flexible and scalable way to incorporate geographic information into pricing models. They overcome the limitations of entity embeddings and spatial smoothing methods and can be used in combination with other modelling techniques. The embeddings are smooth, transferable, and computationally efficient, making them suitable for large-scale applications in insurance analytics. A further advantage of the proposed framework is that it can be trained on modest computational hardware and easily adapted to new regions or additional spatial data sources, making it practical for in-house deployment at insurance companies and other organisations. 

\added{An important consideration for the deployment of spatial embeddings in pricing is fairness. The embeddings are more granular than traditional territorial variables such as postal codes and may therefore pick up socioeconomic patterns not captured at the postal code level alone. When used as rating factors, they could act as a proxy for characteristics that are protected or sensitive. Combining our framework with the literature on fairness in insurance pricing \citep{cote2025fair, jamotton2026energy} is therefore a relevant direction, both to quantify the extent to which spatial embeddings encode protected characteristics and to construct pricing models that use their spatial information while satisfying fairness requirements.}

\added{A limitation of our case studies is the temporal alignment between the outcome data and the spatial views, which were collected at the time of the study. The framework itself is not restricted in this respect, since the embedding model can be retrained per calendar year, or per any other relevant period, with temporally matched imagery and map data. Such data are available, for example, through the historical satellite archives in Google Earth Engine \citep{gorelick2017google} and snapshots of OpenStreetMap reconstructed from edit timestamps. The reported training times show that such periodic retraining is computationally feasible. Extending the framework with an explicit temporal dimension, so that a single model produces embeddings for any location and point in time, is a promising direction for future work.}

This work represents a step toward integrating the principles of foundation models into actuarial science. By demonstrating how multimodal spatial pretraining produces reusable embeddings that enhance predictive accuracy and generalise across space, we show that actuaries can benefit from adopting foundation-model approaches. Similar to the transfer of language and vision models to countless downstream tasks, spatial foundation models can serve as general-purpose representations that can be adapted for pricing, reserving, portfolio analytics, and emerging climate-related risks. While recent global-scale spatial foundation models, such as AlphaEarth \citep{brown2025alphaearthfoundationsembeddingfield}, illustrate what is possible at the planetary scale, our work highlights a complementary direction: lightweight, region-specific spatial embeddings that can be retrained quickly, incorporate organisation-specific spatial views, and capture fine-grained local structure that global models may overlook. Our work enables further exploration of such models, to evaluate how pretrained spatial representations can be transferred across products, regions, and lines of business, and to incorporate domain expertise in shaping foundation models that are both scientifically rigorous and practically relevant for risk management.

\section*{Acknowledgements}

Katrien Antonio gratefully acknowledges funding from the FWO and Fonds De La Recherche Scientifique-FNRS (F.R.S.-FNRS) under the Excellence of Science (EOS) program, project ASTeRISK Research Foundation Flanders [grant number $40007517$]. The authors gratefully acknowledge support from the FWO network W$001021$N. Christopher Blier-Wong acknowledges financial support from the Natural Sciences and Engineering Research Council of Canada (RGPIN-$2025$-$06879$). We also express our gratitude to the editor and reviewers for their valuable feedback, which has greatly enhanced the quality of this article.

\section*{Competing interests}

The authors declare no potential conflicts of interest. 

\section*{Supplemental material}

The code to train the model architecture can be found via the GitHub repository: \url{https://github.com/freekholvoet/MultiviewSpatialEmbeddings}. The trained models, as defined in Table \ref{tab:hyperparam}, are available in \citet{holvoet_mvmodels_2025}.


\bibliographystyle{apalike-doi}
\bibliography{references}

\newpage
\appendix
\section{OpenStreetMap tags per category}\label{app:osmtags}

Table \ref{tab:osm_categories} lists the OpenStreetMap tags used in each of the six categories as defined in Section \ref{sec:spatialdata}. Each category lists the tag type, as defined by OpenStreetMap, along with the names of the tags collected per type.

\begin{table}[ht!]
    \centering
    \begin{NiceTabular}{m{2cm}m{5cm}m{7cm}}[
            code-before = \rowcolor[HTML]{FFFFFF}{1,2,4,6}
            \rowcolor[HTML]{FFFFFF}{3,5,7}
        ]
        \toprule
        \textbf{Category} & \textbf{OpenStreetMap tag type} & \textbf{Tag names} \\
        \noalign{\hrule height 0.3pt}
        shop  & amenity & shops \\
        tourism & amenity & tourism \\
        food\& drink & amenity & bar, cafe, pub, restaurant, fast food, food court, events venue, nightclub \\
        health & amenity & clinic, dentist, doctors, hospital, pharmacy, social facility \\
          & healthcare & clinic, doctor, hospital, pharmacy \\
        education & amenity & college, kindergarten, research institute, school, university \\
        public services & amenity & fire station, police \\
        \bottomrule
    \end{NiceTabular}
    \caption{The OpenStreetMap tags, with the tag type, for each of the six categories defined in Section \ref{sec:spatialdata}.}
    \label{tab:osm_categories}
\end{table}

\section{Hyperparameter selection}\label{app:hypertesting}

Hyperparameter selection for the spatial embedding model was carried out during model construction. As the embeddings are trained in a self-supervised manner, their quality cannot be directly evaluated using standard predictive metrics. Instead, the aim was to identify configurations that produced stable training behaviour and embeddings containing spatially informative structure.

A range of model configurations was tested for the OpenStreetMap encoder, location encoder, and fusion layer, varying the number of filters, neurons, and learning rates. Each configuration was pretrained for a fixed number of epochs and evaluated on two criteria. First, the contrastive loss (CLIP loss) was monitored to ensure stable convergence. Second, the resulting embeddings were applied in small-scale downstream tasks, such as logistic regression for country classification and a regression model for population density prediction. For both downstream tasks, we compared the performance of models with the embedding vector as input versus those with latitude and longitude coordinates. Configurations showing both stable loss behaviour and high downstream classification accuracy were retained.

The final hyperparameter values, reported in Section~\ref{sec:hyperparam}, combine established settings from \citet{klemmer2023satclip} with those identified through these preliminary tests. Smaller convolutional architectures ($8$ and $16$ filters of size $k=1$) and a $128$-neuron fusion layer yielded consistent results, while larger models offered limited improvement and less stable optimisation. As in most self-supervised learning setups, the hyperparameter selection is therefore empirical, guided by training stability and representational quality rather than formal optimisation.

\section{Land use classification codes}\label{app:landuse}

Table \ref{tab:landuse_codes} lists all land use codes in the dataset in Section \ref{sec:fr_eda}. The original description in French is given, along with a translation. 

\begin{table}[ht!]
    \centering
    \begin{NiceTabular}{m{1.5cm}m{5cm}m{6cm}}[
            code-before = \rowcolor[HTML]{FFFFFF}{1,2,4,6,8,10,12,14,16,18,20,22,24,26}
            \rowcolor[HTML]{FFFFFF}{3,5,7,9,11,13,15,17,19,21,23,25,27}
        ]
        \toprule
        \textbf{Code} & \textbf{Original description} & \textbf{English translation} \\
        \noalign{\hrule height 0.3pt}
        AB  & terrains à bâtir      & building plot \\
        AG  & terrains d'agrément   & amenity land \\
        B   & bois                  & woodland \\
        BF  & futaies feuillues     & deciduous forest \\
        BM  & futaies mixtes        & mixed forest \\
        BO  & oseraies              & willow grove \\
        BP  & peupleraies           & poplar grove \\
        BR  & futaies résineuses    & coniferous forest \\
        BS  & taillis sous futaie   & coppice with standards \\
        BT  & taillis simples       & simple coppice \\
        CA  & carrières             & quarry \\
        CH  & chemin de fer         & railway \\
        E   & eaux                  & water body \\
        J   & jardins               & garden plot \\
        L   & landes                & heathland \\
        LB  & landes boisées        & wooded heath \\
        P   & prés                  & meadow \\
        PA  & pâtures               & pasture \\
        PC  & pacages               & grazing land \\
        PE  & prés d'embouche       & grazing meadow \\
        PH  & herbages              & grassland \\
        PP  & prés plantes          & crop meadow \\
        S   & sols                  & residential plot \\
        T   & terres                & arable land \\
        TP  & terres plantées       & planted arable land \\
        VE  & vergers               & orchard \\
        VI  & vignes                & vineyard \\
        \bottomrule
    \end{NiceTabular}
    \caption{Land-use classification codes for the variable \texttt{code\_nature\_culture} in the dataset introduced in Section \ref{sec:fr_eda}. The original French description is provided alongside the English translation.}
    \label{tab:landuse_codes}
\end{table}

\end{document}